\documentclass[aps,prb,twocolumn,preprintnumbers,superscriptaddress,amsmath,floatfix]{revtex4}
\usepackage{graphicx} 

\usepackage{amsmath}
\usepackage{amssymb}
\usepackage{enumitem}
\usepackage{multirow}
\usepackage{physics}
\usepackage[dvipsnames]{xcolor}
\usepackage[T1]{fontenc}
\usepackage{graphicx}
\usepackage{dcolumn}
\usepackage{bm}
\usepackage[retainorgcmds]{IEEEtrantools}
\usepackage[colorlinks=true,linkcolor=blue,citecolor=blue,urlcolor=blue]{hyperref}
\usepackage{bbold}
\usepackage{mathtools}
\usepackage{accents}
\usepackage{verbatim} 
\usepackage{dutchcal}
\usepackage{gensymb}

\renewcommand{\figurename}{\textbf{Fig.}}
\usepackage{xcolor}
\usepackage[normalem]{ulem}
\newcommand{\editor}[2]{%
  \expandafter\newcommand\csname #1note\endcsname[1]{%
    \textcolor{#2}{(\textbf{#1:} \textit{##1})}}%
  \expandafter\newcommand\csname #1\endcsname[1]{%
    \textcolor{#2}{##1}}%
  \expandafter\newcommand\csname #1cancel\endcsname[1]{%
    \textcolor{#2}{\sout{##1}}}%
  \expandafter\newcommand\csname #1change\endcsname[2]{%
    \textcolor{#2}{\sout{##1} ##2}}%
  \newenvironment{#1text}{\color{#2}}{\color{black}}
}
\definecolor{BluBondi}{rgb}{0.00,0.58,0.71}
\definecolor{Orange}{rgb}{1.00,0.45,0.00}
\editor{AG}{BluBondi}
\editor{AGnew}{Orange}

\begin{document}
\title{Ultraviolet optical conductivity, exciton fine-structure and dispersion of freestanding monolayer h-BN}

\title{Exciton dispersion fine structure and deep ultraviolet optical conductivity of freestanding two-dimensional h-BN}

\author{Jinhua Hong}
\affiliation{College of Materials Science and Engineering, Hunan University, Changsha 410082, China}

\author{Alberto Guandalini}
\affiliation{Dipartimento di Fisica, Università di Roma La Sapienza, Piazzale Aldo Moro 5, I-00185 Roma, Italy}

\author{Weibin Wu}
\affiliation{College of Materials Science and Engineering, Hunan University, Changsha 410082, China}

\author{Haiming Sun}
\affiliation{SANKEN (The Institute of Scientific and Industrial Research), The University of Osaka, Mihogaoka 8-1, Osaka, Ibaraki 567-0047, Japan}
\author{Fuwei Wu}
\affiliation{College of Materials Science and Engineering, Hunan University, Changsha 410082, China}
\author{Shulin Chen}
\affiliation{College of Materials Science and Engineering, Hunan University, Changsha 410082, China}
\author{Chao Ma}
\affiliation{College of Materials Science and Engineering, Hunan University, Changsha 410082, China}
\author{Kazu Suenaga}
\affiliation{SANKEN (The Institute of Scientific and Industrial Research), The University of Osaka, Mihogaoka 8-1, Osaka, Ibaraki 567-0047, Japan}
\author{Thomas Pichler}
\affiliation{University of Vienna, Faculty of Physics, Strudlhofgasse 4, A1090 Vienna, Austria}
\author{Francesco Mauri}
\affiliation{Dipartimento di Fisica, Università di Roma La Sapienza, Piazzale Aldo Moro 5, I-00185 Roma, Italy}

\date{\today}
\begin{abstract}
    Excitons govern the light-matter interaction in 2D gapped materials with intrinsically large binding energies.   
    In spite of plentiful optical measurements in the visible for semiconducting transition-metal dichalcogenides,  we still lack optical-absorption studies of the exciton structure of insulating 2D materials that requires UV light. 
    Moreover, measurements of the momentum dispersion of excitons in the vicinity of optical limit are rare owing to low resolutions but hold the key to reveal quasiparticle interactions. To close this gap, we employ high momentum resolution electron energy loss spectroscopy ($q$-EELS) to explore exciton dispersions of mono- and few-layer hexagonal boron nitride. Surprisingly, we reveal a fine structure of the first bright exciton dispersion band composed by two features (A and A$'$), visible only at small momentum, not predicted by Bethe-Salpeter calculations.
    Introducing an optical conductivity approximation (OCA), we extract from the experimental $q$-EELS spectra the ultraviolet (UV) optical conductivity at zero momentum, $\sigma(\omega)$, and discuss the exciton fine structure in $\sigma(\omega)$, consistent with previous photoluminescence observations. Our findings establish a general methodology to probe the fine structure of exciton dispersions, providing new insights into exciton-phonon sidebands and eventually polarons in low-dimensional materials.
\end{abstract}

\maketitle
Excitons play an essential role in the excitation properties of two-dimensional (2D) band insulators~\cite{Mingsheng_2013} subject to external photons, electrons, or neutrons. 
Two-dimensional materials, such as hexagonal boron nitride (h-BN)\cite{Zhang_2017,wang2019epitaxial,chen2020wafer,caldwell2019photonics}, transition metal dichalcogenides (TMDs) \cite{wang2012electronics,PhysRevLett.105.136805} and phosphorene\cite{li2017direct,qiao2014high,chen2020widely,zhou2023pseudospin}, preserve an intrinsic large exciton binding energy due to their weak screening in the atomic-scale thickness dimensionality\cite{jiang2017scaling,ugeda2014giant}.
This gives rise to sharp excitonic peaks followed by absorption continuum\cite{chernikov2014exciton}, typical in 2D excitonic systems.
As an example, h-BN/TMDs/h-BN heterostructures have long been an ideal platform to explore the interlayer\cite{rivera2015observation}, charge-transfer\cite{rosati2023interface}, Moiré excitons\cite{alexeev2019resonantly,jin2019observation,seyler2019signatures}, trions\cite{mak2013tightly} and even higher-order charge complexes, building up a rich landscape of the excitonic physics in 2D semiconductors\cite{regan2022emerging}.
Also, exotic phases of matter such as exciton-mediated superconductivity, exciton condensates or insulators\cite{kogar2017signatures}, superfluidity\cite{mei2025evidence} are found in these semiconductor systems.
However, in contrast with TMDs, the intrinsic exciton band structure of 2D UV-gapped h-BN is seldom explored\cite{elias2019direct,fu2025indirect,rousseau2021monolayer,henriques2019optical}, but vital for the understanding of low-dimensional devices encapsulated by h-BN and for deep-UV optics applications. 
    
In most optical experiments, the excitonic behavior involves mostly vertical electronic transition preserving negligible momentum transfer ($q \sim 2\pi\theta/\lambda \sim 10^{-4} $\AA$^{-1}$).
Thus, the degree of freedom of momentum is seldom involved, hindering a thorough measurement of the many-body interactions such as the exciton band structures\cite{onida2002electronic,qiu2013optical} and their couplings with finite momentum phonons/photons. 
To access non-negligible momentum, inelastic X-ray/neutron scattering or electron energy loss spectroscopy (EELS) is required. For the former X-ray/neutron techniques, the experimental challenges lie in either the large required crystal size (mm), difficult to achieve in practice, or the need of substrates causing a strong quenching or decay of the excitonic peaks. 
EELS in a transmission electron microscope (TEM) is a feasible way to measure exciton band structures of h-BN sheets, as it directly probes the longitudinal dielectric response resolved both in energy and momentum~\cite{Egerton_book,Ibach_book} and may be applied to nanometer-scale and suspended samples  \cite{gaufres2019momentum,schuster2015anisotropic,hong2020probing}, provided that $q$ resolution is sufficiently high to distinguish different $q$ signals.

In theoretical calculations, exciton band structures and related properties are predicted starting from a GW band structure~\cite{Hedin_1965,Strinati_1982,Hybertsen_1986,Godby_1988} and solving the Bethe-Salpeter equation (BSE)~\cite{Hedin_1965,Strinati_1988,Onida_2002}, eventually at finite momenta~\cite{Gatti_2013}. In particular, several works have investigated the excitonic band structure of h-BN\cite{cudazzo2016exciton} and TMDs\cite{qiu2015nonanalyticity,fugallo2021exciton}, finding a double-degenerate optical exciton, one bright and the other dark, with linear and parabolic dispersions respectively.
Excitonic effects have even been studied in materials without bounded excitons, such as graphene~\cite{guandalini_2023,Guandalini_phBSE}.
However, lattice vibrations and their interactions with excitons are usually neglected within the GW-BSE framework. Some pioneering works with included exciton-phonon coupling~\cite{Antonious_2022,Lechifflart_2023,Marini_2024} or exciton-polaron\cite{dai2024theory,sio2023polarons}, executed on top of frozen nuclei GW-BSE, demonstrated the importance of mixed exciton-phonon excitations in the description of the h-BN photoluminescence spectra.
However, controversial results are obtained both in experiments~\cite{Elias_2019,Wang_2022,Shima_2024}, probably due to screening effects of the substrate, and in theoretical simulations~\cite{Lechifflart_2023,Marini_2024}, due to the different approximations involved.\\

In this work, we employ momentum ($q$) resolved EELS ($q$-EELS) in a TEM and beyond-GW-BSE calculations to study the exciton dispersions of mono- and few-layer h-BN, fully characterizing this UV-range excitonic system. 
We find a fine-structure composition of the lower-linear dispersive bright exciton (A-A$'$) with a splitting gap of $\Delta \sim 0.25$ eV and a linear dispersive branch (B).
The A-A$'$ splitting is not present in GW-BSE calculations, while the weighted average of the A-A$'$ fine structure and B exciton dispersions are instead in accordance with the calculated exciton dispersions.
We ascribe the A$'$ (or A) absorption to be dominated by exciton-phonon coupling, as not found in frozen lattice calculations and already qualitatively described in previous works\cite{marini2008ab}.\\

Using an optical conductivity approximation (OCA) to model the $q$-EEL spectra, we are able to extract the optical conductivity $\sigma(\omega)$ and characterize the A$'$ satellite at $q=0$. We point out the OCA extraction provides the experimental optical conductivity in the deep-UV regime, which is basically impossible to get accurately by UV optics especially for freestanding mono and few-layer materials. Finally, by checking the thickness dependence of the exciton fine structure, we find A$'$ satellite is manifested in the monolayer and becomes more and more invisible concomitant to the decreasing A-A$'$ gap as the thickness increases. This indicates the exciton-phonon coupling effect is strongest in the monolayer limit for the reshaping of phonon satellites of excitons. Our work offers a successful methodology to investigate the complicated exciton band structure in 2D materials with considerable exciton-phonon interaction, which can be generalized to other low-dimensional systems.\\
\\
\noindent\textbf{Experimental exciton dispersion fine structure}\\
\begin{figure}[h!] 
    \includegraphics[width=\textwidth/2, keepaspectratio]{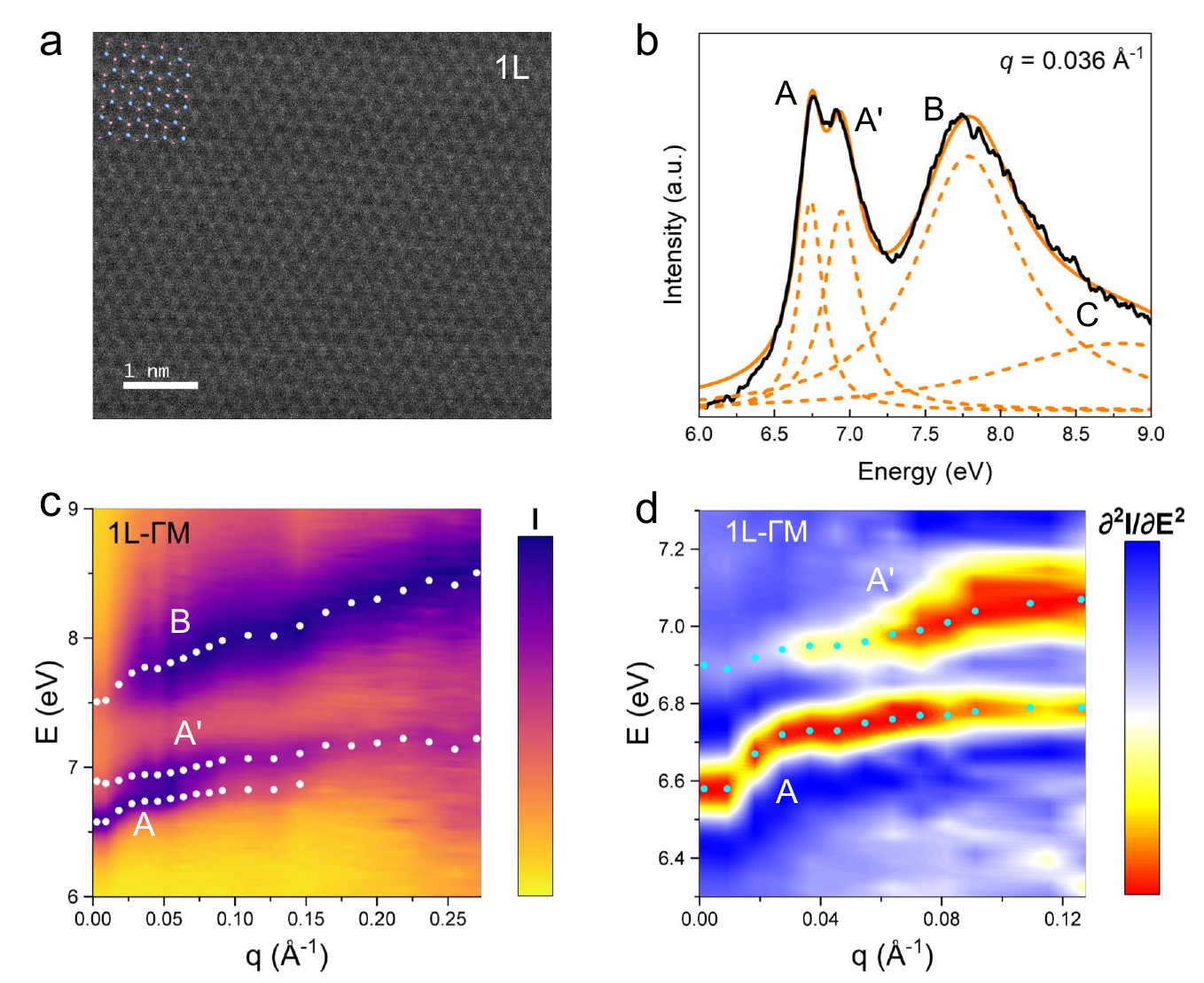}
    \caption{\textbf{The q-E diagrams of monolayer h-BN by $q$-EELS.} (a): Atomically resolved ADF-STEM image of monolayer h-BN. Scale bar: $1$ nm. Pink balls are N atoms and blue ones correspond to B atoms. (b): EEL spectrum at a finite momentum ($q=0.036$ \AA$^{-1}$) in black. Peak positions and area are identified via a Four Lorentzian fitting model where each Lorentzian is plotted with dashed orange lines. The continuous orange line indicates the sum of the four Lorentzians, smoothing the experimental signal.
    (c): Experimental $q$-$E$ diagram for the valence electron energy-loss (EEL) spectra in monolayer h-BN along the $\Gamma$M direction.
    (d): Second order derivative of the spectrum intensity of A, A$'$ excitations. White (c) and cyan (d) dots correspond to the peak positions at each $q$. }
    \label{fig:exp}
\end{figure}
The experimental EEL spectra are collected in the diffraction mode of a monochromated TEM at 60 kV with an energy resolution of $\Delta E = 45$ meV and momentum resolution of $\Delta q = 0.02$ \AA$^{-1}$. Fig. \ref{fig:exp}a shows atomically resolved annular dark field scanning transmission electron microscopy (ADF-STEM) image of high-quality monolayer h-BN (see more in Figs. S1-S3), with the inset superimposed hexagonal atomic model.
Through the annealing in vacuum at $300$ $^{\circ}$C for 6h, the h-BN surface becomes clean and carbon contamination is prevented to make sure EEL spectra reflect the intrinsic excitonic properties of monolayer h-BN (Figs. S4-S8).

Then, we obtain the $q$-EEL spectra along $\Gamma$M and $\Gamma$K direction.
Fig. \ref{fig:exp}b shows in black an example of the spectrum at a fixed momentum ($q=0.036$ \AA$^{-1}$) along the $\Gamma$M direction.
We identify two excitonic branches: one composed by two peaks (A, A$'$) at $\sim 6.5\text{-}7$ eV and another composed of a single asymmetric peak (B) at around $\sim 8$ eV with a longer high-energy tail.
To identify the maxima and oscillator strengths of all the excitations, we fit the EEL spectra with a four Lorentzian model (4LM, more details in Fig. S9-S10), which accurately reproduce the raw experimental data. 

\begin{figure}[t] 
    \includegraphics[width=0.5\textwidth,keepaspectratio]{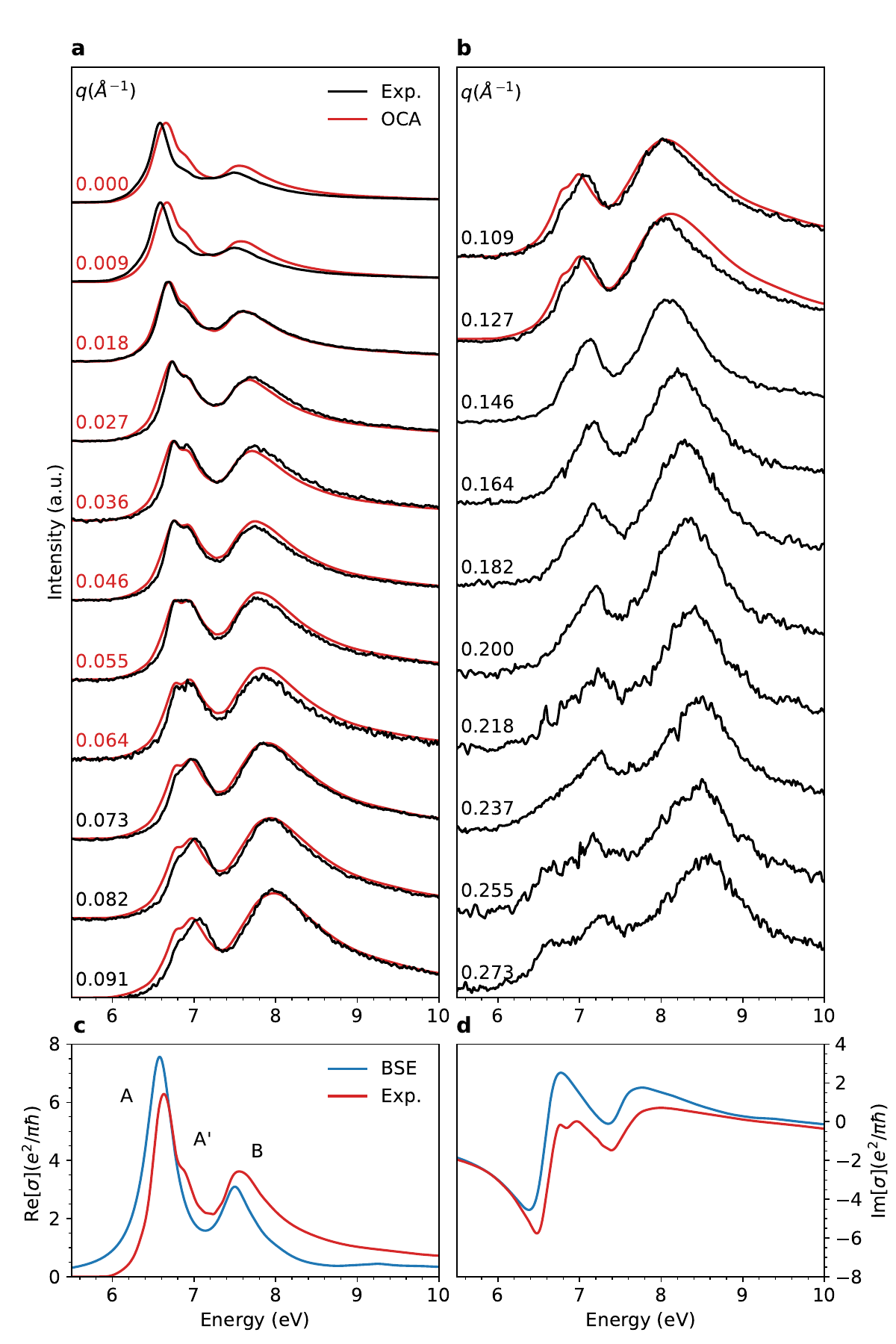}
    \caption{\textbf{Exciton structure of monolayers in the low q range and optical conductivity.} 
    (a)-(b): $q$-EELS experimental spectra (black) and obtained from the optical conductivity approximation [red lines of panels (c) and (d)]. Red momentum transfers correspond to the data used to extract the optical conductivity. We note the optical conductivity approximation slowly deteriorates over the increasing $q$.    
    Real (c) and imaginary (d) parts of the 2D optical conductivity in units of the quantum conductance ($G_0 = e^2/\pi\hbar$) are extracted from experiments through the optical conductivity approximation (red) and calculated with the Bethe-Salpeter equation at fixed nuclei (blue).
    The presence of the A$'$ peak in the experimental optical conductivity while missing in the BSE one is an indication of the phonon nature of the A$'$ excitation.
}
    \label{fig:spectra}
\end{figure}

Fig. \ref{fig:exp}c shows the q-E diagram of the low-energy electron excitations of monolayer h-BN within limited $q < 0.3$ \AA$^{-1}$ along the $\Gamma$M direction.
Due to the detection sensitivity and deceasing signal-to-noise ratio, the exciton peaks can only survive within $1/5$ of the Brillouin zone nearby the $\Gamma$ point ($|\Gamma M|=1.45$ \AA$^{-1}$, $|\Gamma K|=1.67$ \AA$^{-1}$).
Within this range, the in-plane $\Gamma$M/$\Gamma$K anisotropy can be ignored (Extended Data Fig.~\ref{fig:S17}), since the dispersion structure shows almost no difference. In Fig. \ref{fig:exp}c, the white dots correspond to the maxima found by the 4LM fitting of the spectra normalized to their maximum at each $q$ (see Figs. \ref{fig:spectra}c-d ). The lower A exciton band starts at $6.6$ eV at $q=0$, a little higher than the reported optical measurements\cite{elias2019direct,henriques2019optical,rousseau2021monolayer,fu2025indirect}.
Here small deviations in the absolute energies may be attributed to different substrates, or, in our case, the missing of substrates.
The sideband A$'$ appears and gradually dominates the fine structure as $q$ shifts away from the optical limit $q=0$, while the lower A exciton is observed to survive within low q < $0.15$ \AA$^{-1}$.
Therefore, a high momentum resolution is very key to detecting this exciton fine structure due to its low-momentum regime, not found previously\cite{liu2025direct}.
Compared with A-A$'$ fine structure, the upper B exciton shows instead an almost linear dispersion with increasing relative intensity as $q$ increases.

To better visualize the A-A$'$ splitting, second-order derivatives $\partial^2 I/\partial E^2$ of the EEL intensity (I) are plotted in Fig. \ref{fig:exp}d, a widely used mathematical treatment in optical spectroscopy.
We note the peak positions by 4LM in Fig. \ref{fig:exp}c are in agreement with $\partial^2 I/\partial E^2$ minima (cyan dots) in all the $q$ range.
Here, a linear dispersion for A$'$ and fractional-polynomial dispersion for A can be seen clearly within the vicinity of the optical limit (Extended Data Fig.~\ref{fig:S17}e-f).
The less dispersive band A$'$ is located above the fractional-polynomial branch A with a small splitting gap $\Delta \sim 0.25$ eV.\\ 

\noindent
\textbf{q-EEL spectra and extraction of the deep-UV optical conductivity}\\
In Figs. \ref{fig:spectra}a-b, we show the experimental EEL spectra in black at increasing $q$ of monolayer h-BN. One may note the $q=0.009$ \AA$^{-1}$ and the $q=0$ spectra are basically identical because they fall within experimental resolution which is also averaging the higher $q$ values. All peaks present a blue shifting as $q$ increases along $\Gamma$M  (see also $\Gamma$K direction and multi-layer in Extended Data Fig.~\ref{fig:spectra_extended} and larger energy range $\sim 20$ eV shown in Fig. S11-S12). The lowest exciton A shows a decreasing oscillator strength as $q$ increases, while the excitons A$'$ and B shows obvious increasing intensity as $q$ shifts away from $\Gamma$ point.
The splitting of the A-A$'$ peak, for $q<0.15$ \AA$^{-1}$, is clearly visible.
For $q>0.15$ \AA$^{-1}$, the two features merge into an asymmetric unique shape.

Within the optical conductivity approximation (OCA), we assume that low-$q$ EEL spectra are fully determined by the complex optical conductivity $\sigma(\omega)$ at $q=0$ (see the Method section and Extended Data Figs.~\ref{fig:OCA_range},\ref{fig:fit_explain}) and that the $q$-dependence of position, shape and intensity of the spectral features are only due to the long-range component of the Coulomb interaction.
Taking advantage of this hypothesis, we have been able to extract from the low-$q$ EEL spectra ($q<0.07$ \AA$^{-1}$) the deep-UV real and imaginary parts of the optical conductivity, which is almost inaccessible from UV optics. Then, we compare it with the GW-BSE one in Figs. \ref{fig:spectra}c-d respectively (see also Fig. S13).
We verify the accuracy of the extracted $\sigma(\omega)$ by comparing the OCA calculated spectra with the experimental $q$-EEL data for $q<0.13$ \AA$^{-1}$ in Figs. \ref{fig:spectra}a-b with a remarkable accuracy, in line with the OCA accuracy for GW-BSE calculations in the Extended Data Fig.\ref{fig:OCA_range}.

The A$'$ peak, not shown in the GW-BSE calculation, is present also in $\sigma(\omega)$, in analogy with Ref.~\onlinecite{Marini_2024}, where exciton-phonon coupling assisted absorption is included in the GW-BSE absorption spectrum via a cumulant expansion of the exciton-phonon self-energy.
We can therefore reasonably conclude that the A$'$ peak originates from the combined creation of an exciton and a phonon both at the Brillouin zone corner K, as previously claimed to happen in photoluminescence~\cite{Marini_2024}.
We stress that the $\sigma(\omega)$ has been extracted from the experiments with the only assumption that the conductivity is not dispersive in the vicinity of optical limit.
Thus, it may be used as a benchmark for future calculations trying to reproduce exciton-phonon coupling effects in optics.\\
\\
\noindent\textbf{Quantitative analysis of the optical and finite momentum exciton properties}\\
We extract energies and oscillator strengths of the A, A$'$ and B peaks in the optical limit $q=0$ from the $\sigma(\omega)$ by OCA and $q$-EEL spectra by 4LM fit, summarized in Tab.~\ref{tab:sigma_parameters}. The $\sigma$ analysis by OCA agrees quite well with the EELS results for both peak positions and intensities, which also persists into bilayer case.
Theoretically, the $q=0$ EEL spectra divided by $\omega$ is proportional to $\sigma(\omega)$ apart for an energy-independent factor~\cite{Guandalini_mes_nothing}.
However, in practice the $q=0$ spectra include also contributions from finite momentum signals due to the finite resolution of $q$, causing small discrepancies with $\sigma(\omega)$.
\begin{table}
\centering
\begin{ruledtabular}
\begin{tabular}{ccccc|ccc}
&\multicolumn{4}{c}{$\sigma$ analysis}&\multicolumn{2}{c}{EELS $q=0$}\\
& $1\Gamma$M   & $1\Gamma$K & 2$\Gamma$M & 1BSE& $1\Gamma$M   & $1\Gamma$K & 2$\Gamma$M\\
\hline
$E_{\mathrm{A}}$ (eV)&6.62&6.62&6.55&6.57&6.58&6.58&6.55\\
$E_{\mathrm{A'}}$ (eV)&6.87&6.88&6.75&-&6.89&6.91&6.79\\
$E_{\mathrm{B}}$ (eV)&7.60&7.57&7.50&7.53&7.51&7.48&7.43\\
$E_{\mathrm{A'}}$-$E_{\mathrm{A}}$ (eV)&0.25&0.26&0.20&-&0.31&0.33&0.24\\
$I_{\mathrm{A'}}/I_{\mathrm{A}}$&0.37&0.34&0.65&-&0.23&0.28&0.39\\
$(I_{\mathrm{A}}+I_{\mathrm{A'}})/I_{\mathrm{B}}$&0.99&0.84&0.66&2.65&1.53&1.49&1.03\\
$\Sigma/N$ &1.15&1.12&1.10&1.10&-&-&-\\
$r_{\mathrm{eff}}/N$(\AA)&5.80&6.07&6.49&5.42&-&-&-\\
\end{tabular}
\end{ruledtabular}
\caption{Peak position and intensity ratio of the optical conductivity $\sigma(\omega)$ extracted via OCA and experimental $q=0$ EELS cross section divided by $\omega$ for mono- (1) and bi- (2) layer. 
$N$ is the number of layers. $\Sigma = 2A\int_{0}^{E_{\mathrm{f}}}\sigma(E)dE/\pi$ with $E_{\mathrm{f}} = 10$ eV is the sum rule of $\sigma(\omega)$. 
The effective radius $r_{\mathrm{eff}}$ is the first order expansion of the static dielectric constant with respect to $q$, so implicitly defined as:
$\epsilon(q,\omega=0) \approx 1+r_{\mathrm{eff}}q$. See Refs.~\onlinecite{Sohier_2016,Sohier_2017} for more details about the definition of the effective radius.}\label{tab:sigma_parameters}
\end{table}

\begin{figure}[t] 
    \includegraphics[width=0.5\textwidth, keepaspectratio]{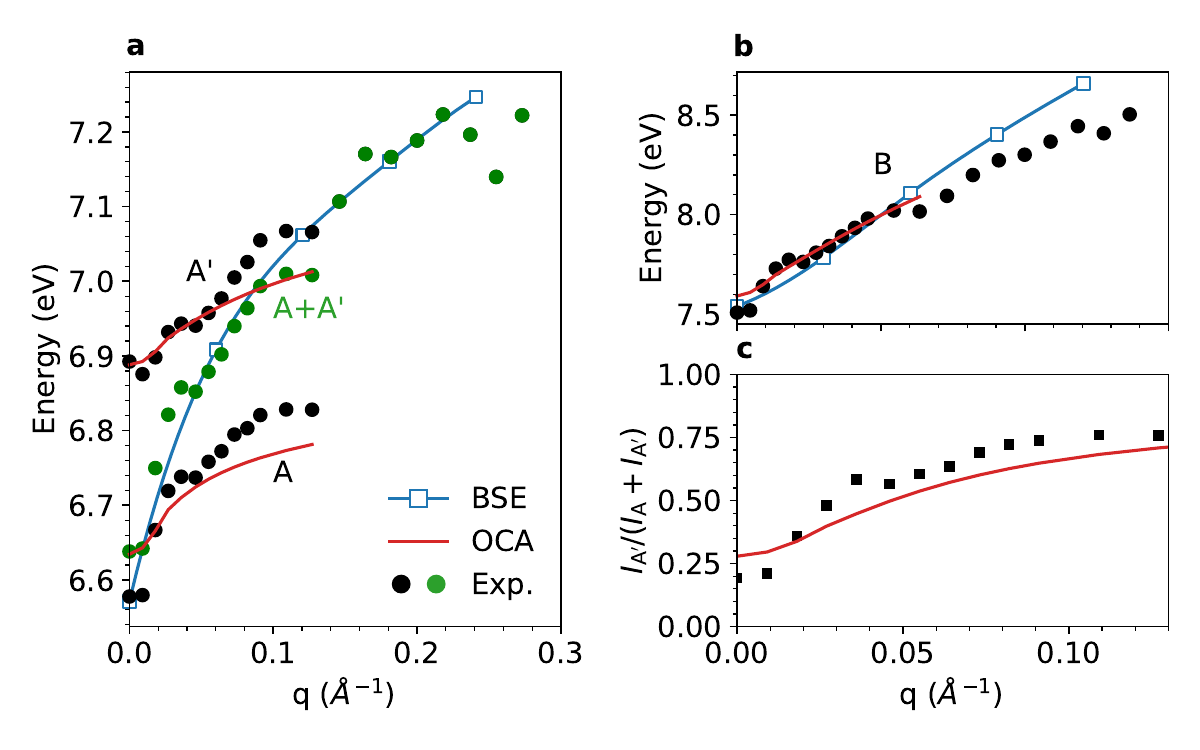}
    \caption{\textbf{Energy and oscillator strength dispersions of monolayer h-BN along $\Gamma$M direction.} (a): Exciton dispersions of the A and A$'$ peaks in black and its weighted average (($E_AI_A+E_{A'}I_{A'})/(I_A+I_{A'})$) in green, and OCA (red) and BSE (blue) calculated dispersions. The OCA has been restricted to its range of validity up to $q_{\mathrm{max}} \approx 0.12$ \AA$^{-1}$
    (b): The dispersion of the B peak obtained by experiment (black), OCA (red) and BSE (blue).
    (c): Fraction of the intensity of A$'$ peak with respect to the A+A$'$ peaks.
    The same analysis is performed along the $\Gamma$K direction and for bi- and three-layer along the $\Gamma$M direction. Results can be found in the Extended Data Fig.~\ref{fig:dispersion_extended}.
    }
    \label{fig:dispersion}
\end{figure}

Extending our methods to all $q$ range (Fig. S9-S10), we plot the experimental dispersions for splitting excitons A and A$'$ as black dots, in comparison with the calculated curves in Fig.~\ref{fig:dispersion}a.
Compared with the less dispersive peak A$'$, the lower exciton A presents a more obvious fractional-polynomial relation, in contrast with the parabolic dispersions in TMDs\cite{hong2020probing}. According to our GW-BSE calculation, and previously discussed in the literature~\cite{Galvani_2016}, the exciton A is degenerate with a parabolic dark exciton band at $\Gamma$ point (Extended Data Fig.~\ref{fig:BSE_dark}), arising from the double degeneracy of K and K$'$ valleys of the electronic structure.
Nevertheless, the parabolic band is not observed in the experimental spectra, owing to its dipole forbidden rule predicted by GW-BSE calculations.
The OCA predicted (in red curves) A-A$'$ fine structure in Fig.~\ref{fig:dispersion}a shows a good accuracy with respect to the experimental dispersions, validating \textit{a posteriori} the extraction procedure.
Similar splitting fine structure can be found in $\Gamma$K direction in Extended Data Fig. ~\ref{fig:dispersion_extended}.

Given a coarse accuracy or a lower energy resolution, we obtain a weighted-average exciton dispersion for the overall A+A$'$ structure, ($E_AI_A+E_{A'}I_{A'})/(I_A+I_{A'})$, marked by the green dots in Fig.~\ref{fig:dispersion}a.
The A+A$'$ averaged dispersion is in good accuracy with GW-BSE calculations, which is taken as a linear dispersion by Liu et al\cite{liu2025direct}.
We thus conclude the linear exciton dispersion is consistent with the frozen nuclei approximation, while exciton-phonon coupling creates satellite peaks.

For the high energy peak B, its dispersion in Fig.~\ref{fig:dispersion}b presents an almost linear relation, coinciding well with the GW-BSE calculated dispersions. Compared with exciton A, the linewidth of B is much larger, indicating the complicated exciton components. This is confirmed by GW-BSE calculations~\cite{Galvani_2016}, which predicts the B peak to consist of several excitons with similar energies. 

 Within the A+A$'$ structure, we plot the ratio of the intensity of A$'$ to that of A+A$'$, namely $I_{\mathrm{A'}}/(I_{\mathrm{A}}+I_{\mathrm{A'}})$, as a function of $q$ (Fig.~\ref{fig:dispersion}c). We see an obvious increase behavior, that is, the sideband A$'$ plays a more and more dominant role in the exciton lineshape, indicating a more explicit phonon satellite as $q$ shifts away from the optical limit.\\
\begin{figure}[t] 
    \includegraphics[width=0.5\textwidth, keepaspectratio]{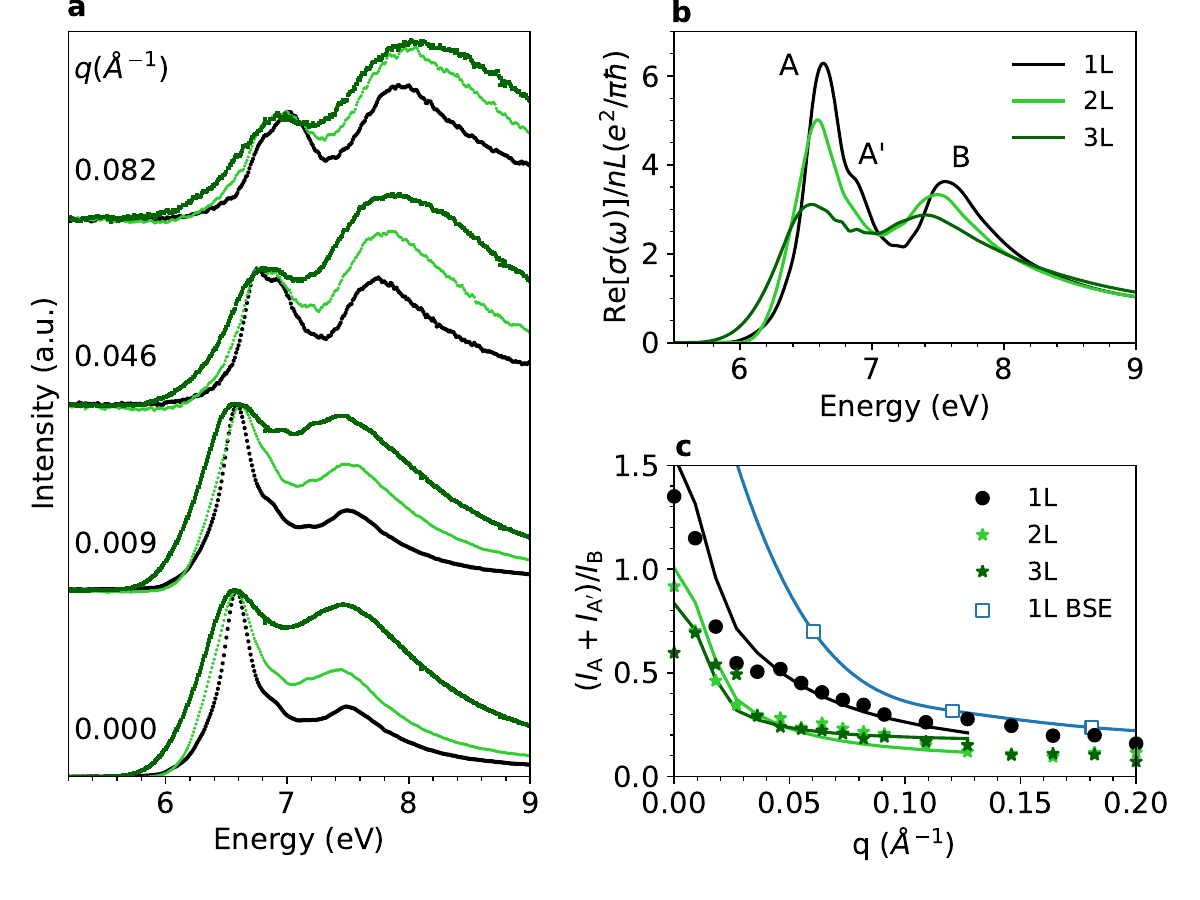}
    \caption{\textbf{Thickness sensitivity of the visualization of exciton fine structure.} (a): A comparison of the $q$-EEL spectra of mono- and few-layer h-BN. (b): Real part of the optical conductivity of mono- (black), bi- (light green) and tri- (dark green) layer of h-BN.
    (c): Intensity ratio of the overall A+A$'$ peak with respect to the B peak from mono to tri-layer.
    As a reference, the intensity ratio of the monolayer calculated by GW-BSE is shown in blue.
    }
    \label{fig:mono_vs_multilayer}
\end{figure}

\noindent\textbf{Thickness dependent phonon satellite }\\
We finally discuss the excitonic properties of h-BN at increasing number of layers.
Previously reported photoluminescence experiments~\cite{Elias_2019,Wang_2022,Shima_2024} show different spectra at different environments, indicating that environmental effects considerably modify the excitonic properties of monolayer h-BN.
In Fig.~\ref{fig:mono_vs_multilayer}a, we show some selected $q$-EEL spectra of mono-, bi- and three-layer h-BN.
Both mono- and multi-layer h-BN show an obvious weight transfer from A+A$'$ excitons to the B excitons, as seen in Fig.~\ref{fig:mono_vs_multilayer}a qualitatively and in Fig.~\ref{fig:mono_vs_multilayer}c quantitavely.
Here, GW-BSE calculations are not able to predict an accurate low-momentum intensity ratio of the monolayer h-BN.
Increasing the number of layers, the peaks broaden due to the interaction with excitons localized in single layers~\cite{Paleari_2018}, and the intensity ratio $(I_A+I_{A'})/I_B$ decreases at low momenta, which used to act as a fingerprint of the thickness identification~\cite{Nerl_2017}. 
The OCA correctly predicts this trend, indicating that it arises from the 2D dielectric properties of few-layer h-BN, dictated by the 2D Coulomb interaction.

Regarding the fine structure, in Fig.~\ref{fig:mono_vs_multilayer}b we show the extracted $\sigma(\omega)$ of mono-, bi- and tri-layer, whose accuracy is demonstrated \textit{a posteriori} from the ability of OCA $q$-EEL spectra to reproduce experimental spectra (see also Extended Data Fig.~\ref{fig:spectra_extended}).
As found from the EEL spectra, the phonon satellite peak A$'$ is clearly manifested in the monolayer limit (black curve in Fig.~\ref{fig:mono_vs_multilayer}b). Unfortunately, such kind of fine structure becomes more and more invisible in the $\sigma(\omega)$ broadening probably due to Davydov splitting~\cite{Paleari_2018} at increasing thicknesses.\\

\noindent\textbf{Discussion and outlook}\\
Looking over the exciton fine structure, the splitting gap $ \Delta $ also shows a thickness dependence: decreasing from $0.25$ eV for monolayer, $0.20$ eV for bilayer, then invisible for trilayer (Extended Data Fig.~\ref{fig:S14},~\ref{fig:second_derivative}). The narrowing splitting gap, together with broadening effect, make the sideband A$'$ submerge quickly as the thickness increases (Extended Data Fig.~\ref{fig:monomap}). 
This indicates the A-A$'$ pair has more almost-degenerate exciton structures causing the broadening and invisibility of the phonon satellite at increasing thicknesses. 

In summary, we employed $q$-EELS and beyond-GW-BSE calculations to study the fine structure of the exciton bands of freestanding h-BN at vanishing and finite $q$.
We find the lower bright exciton is composed by two excitations, not found in GW-BSE calculation, while their weighted average is consistent with GW-BSE predictions of linear dispersion.
Separately, the A and A$'$ dispersion show instead a nonlinear dispersion due to their interaction at finite momenta. With the OCA,  we extracted the optical conductivity $\sigma(\omega)$ of mono-, bi- and tri-layer.
We identified the satellite peak A$'$ as due to exciton-phonon coupling, in analogy with photoluminescence experiments and previous calculations with electron-phonon coupling included~\cite{Marini_2024}.
This finding may provide critical insights into exciton manipulation in the vicinity of optical limit, and our methods also provide a route to study  many-body effects such as exciton-polaron and electron-phonon coupling\cite{qu2025mode} in low-dimensional systems.


\noindent\textbf{Methods}\\
\textit{Sample preparation and transfer}\\
High-quality h-BN monolayers (provided by the group of Dr. Wang\cite{wang2019epitaxial}) are grown on copper (110) foil by chemical vapor deposition through ammonia borane reduced under $Ar/H_2$ gas flow. Continuous single-crystal monolayer h-BN is then transferred onto TEM quantifoil grid using wet chemistry method. First, polycarbonate (PC) solution (2\% in $CHCl_3$) is first dropped onto the monolayer sample using spin coating of 2000 r/min, and then get dried in air for 10 min. Then a PC film is formed on top of h-BN with strong van der Waals interaction. Second, the sample is placed on the surface of $H_2O_2:HCl=1:3$ solution to dissolve the 25$\mu$m-thick Cu substrate within 30 min. Then PC film with h-BN thin layers get floated and then are moved into ultrapure water to wash away possible ions on the sample surface. Third, TEM grid is used to pick up the floating sample with the h-BN side closely contacting the holey carbon film of the grid. After natural drying in air, the TEM grid is immersed into $CHCl_3$ to remove the PC assisted layer, and then into acetone and isopropanol to remove possible organic residues on h-BN surface. Finally, the TEM grid is placed into a CVD furnace at low pressures and then annealed at $300 ^{\circ}$C for 6h to remove most surface contaminations. Before the TEM characterization, the samples are also annealed in the chamber at $300 ^{\circ}$C to prevent carbon deposition. \\
\\
\textit{$q$-EELS measurements}\\
All the $q$-EEL spectra are acquired in diffraction mode using parallel beam in a TEM (JEOL Triple C2) at $60$ kV. The TEM is equipped with a double Wien filter monochromator, providing an electron beam with energy resolution of $40\sim 50$ meV. The smallest spectrometer entrance aperture in a diameter of 1 mm is used at a camera length of 120 cm, to achieve a momentum resolution of $\sim 0.02$ \AA$^{-1}$. The q-serial valence spectra are collected along a specific in-plane orientation $\Gamma$M or $\Gamma$K. In the spectrometer, a dispersion of 0.01 eV/channel is set for the spectrum acquisition. Dual EELS mode is used to exclude the zero-loss peak (ZLP) shift, then dwell time of 1 $\mu$s $\sim 0.1$ s is used for the ZLP, and $0.1$ s $\sim 10$ s is used for the valence spectra of interest. Also, time series method eliminates the impact of ZLP vibration, especially for the high q range taking $\sim 40$ min for each spectrum. All the EEL spectra are calibrated using the ZLP and the flat background is removed to normalize the acquired low loss spectra and to do quantitative analysis.\\
\\
\textit{ADF-STEM imaging}\\
Atomically resolved ADF-STEM imaging is achieved on an aberration corrected TEM (JEOL Triple C3) equipped with a cold field emission gun at $60$ kV. A probe current of $20$ pA and acquisition time per pixel of $60$ $\mu$s is used for imaging. For ADF-STEM imaging, the convergence angle is set to $\sim 35$ mrad, and acceptance angle is $70\sim 200$ mard.\\
\\
\textit{GW-BSE calculation}\\
DFT calculations were performed using a plane wave basis set as implemented in the {\sc Quantum ESPRESSO} package~\cite{QE_2020}, with the PBE approximation~\cite{Perdew_1996} and adopting a kinetic energy cutoff of $60$ Ry for the wavefunctions and norm-conserving pseudopotentials for electron-ion interaction.
The {\sc Yambo} code~\cite{yambo_2009,yambo_2019} has been adopted
to compute the quasi-particle band structure,  within the $G_0W_0$ approximation, and the BSE to obtain zero and finite-momentum EEL spectra and the optical conductivity of monolayer h-BN.
The adopted supercell includes a vacuum region along the perpendicular direction of 7 \AA, and a 2D slab Coulomb  cutoff~\cite{Beigi_2006,Rozzi_2006} has been used to avoid spurious interactions.
For what concerns the calculation of the screening interaction $W$, we used a cutoff of $5$ Ry for the size of the dielectric matrix, including up to $400$ states in the sum-over-state of the response function.
For GW calculations, we adopted
the plasmon-pole approximation in the Godby-Needs scheme~\citep{Godby_1989}.
The quasi-particle Dyson equation has been solved with a linearization of the $G_0W_0$ self energy, and we included $400$ states in the sum-over-state of the non-interacting Green's function.
In the solution of the BSE, we included the lower energetic $10$ bands ($4$ valence and $6$ conduction) and an artificial broadening of $250$ meV. 
We carefully verified that a $48\times 48$ Monkhorst-Pack grid converges both the DFT, GW and BSE calculations.
To fix the first bright exciton with the first experimental peak position, we blue-shifted all GW-BSE EEL spectra by $1.32$ eV.\\
\\
\textit{The optical conductivity approximation (OCA)}\\
All quantities in this section are expressed in atomic units, unless stated otherwise.
The EEL cross section of a 2D system is proportional to the imaginary part of the Fourier transform of the density-density response function $\chi\equiv\delta \rho^{(1)}/\delta\phi^{ext}$, with $\phi^{ext}$ the perturbing scalar potential and $ \rho^{(1)}$ the induced charge density, through the following expression derived elsewhere~\cite{Guandalini_mes_nothing}:
\begin{equation}\label{eq:cross_section}
\frac{d^2S}{d\Omega d\hbar\omega}(q,\omega) =
\frac{4A^{u.c.}}{\pi}\frac{1}{(q^2+q_z^2)^2}
\mathrm{Im} \left[-
\chi^{\mathrm{2D}}(q,\omega)\right],    
\end{equation}
where $A^{u.c.}$ is the area of the 2D unit cell,
$q_z = k_i-\sqrt{k_i^2-q^2-2\omega}$ is the momentum loss along the direction of the electron beam~\cite{Guandalini_mes_nothing} and
$k_i$ is the momentum of the incident electron.
We note the Fourier transform is along the 2D periodic directions of the system in terms of the in-plane component of the momentum transfer $q$.
We note both the cross section $d^2S/d\Omega d\hbar\omega$ and $\chi$ depend on the modulus of $q$, as we focus on low momenta around $\Gamma$ for 2D system isotropic along the periodic directions like mono and few-layer h-BN. The extension to anisotropic cases is straightforward and out of the scope of this work.

Via trivial dielectric considerations, that will be the object of a future publication~\cite{Guandalini_to_be_done}, the Fourier transform of the 2D density-density response function $\chi$ can be expressed as
\begin{equation}\label{eq:sigma2chi}
    \chi^{\mathrm{2D}}(q,\omega) = \frac{\sigma(q,\omega)q^2}{i\omega-v^{\mathrm{2D}}(q)\sigma(q,\omega)q^2} \ ,
\end{equation}
where $v^{\mathrm{2D}}(q) = 2\pi /q$ is the 2D Fourier transform of the Coulomb interaction and $\sigma(q,\omega) = \delta j^{\mathrm{2D},(1)}(q,\omega)/\delta E^{\mathrm{tot}}(q,\omega)$ is the momentum dependent conductivity (see for example Ref.~\onlinecite{Nazarov_2015} for more information). $j^{\mathrm{2D},(1)}(q,\omega)$ and $E^{\mathrm{tot}}(q,\omega)$ are the Fourier transforms of the first-order current density and macroscopic total electric field respectively.

As already noticed in the literature~\cite{Guandalini_mes_nothing}, and tested \textit{a posteriori} in the supporting information, $\sigma(q,\omega)$ is a smooth function of $q$, as opposite to $\chi$, due to the diverging, at small $q$, 2D macroscopic Coulomb interaction $v^{\mathrm{2D}}$.
Such 2D dielectric features are responsible for the 2D LO/TO phonon splitting~\cite{Sohier_2017} and the linear/parabolic excitonic dispersion of the bright/dark lower excitonic branch in TMDs~\cite{Qiu_2015}.
Guided by this ansatz, we approximate the finite momentum conductivity with its optical counterpart, thus $\sigma(q,\omega) \approx \sigma(\omega)$, in Eq.~\eqref{eq:sigma2chi}, and find
\begin{equation}\label{eq:OCA}
    \chi^{\mathrm{2D}}(q,\omega) \approx \frac{\sigma(\omega)q^2}{i\omega-v^{\mathrm{2D}}(q)\sigma(\omega)q^2} \ .
\end{equation}
By substituting Eq.~\eqref{eq:OCA} into Eq.~\eqref{eq:cross_section}, so within the optical-conductivity approximation (OCA), there is a one-to-one correspondence between low-momenta EEL spectra and the optical conductivity.
The OCA can thus be used to extract the deep-UV range of the optical conductivity, following the fitting procedure described in the Extended Data Fig.~\ref{fig:fit_explain}.
We note $\sigma(\omega)$ is in general a complex quantity, and the EEL spectra depend on both $\mathrm{Re}[\sigma]$ and $\mathrm{Im}[\sigma]$ through the denominator in the right side of Eq.~\eqref{eq:OCA}.
The OCA is exact at $q=0$, and there is always at least a vicinity around $q=0$ for which it is justified.
For the specific case of monolayer h-BN, we tested the range of validity of the OCA by comparing \textit{ab-initio} GW-BSE calculations obtained by Eq.~\eqref{eq:sigma2chi} with OCA calculations, obtained via Eq.~\eqref{eq:OCA}, starting from a GW-BSE optical conductivity.
We find the OCA is very accurate for monolayer h-BN for $q<0.07$ \AA$^{-1}$
Results can be found in the Extended Data Fig.~\ref{fig:OCA_range}.\\
\\
\textbf{Data availability}\\
All related data analyzed or processed in the manuscript are available from the corresponding authors on reasonable request.\\ 
\\
\textbf{Code availability}\\
The codes for the theoretical calculation are available upon request.\\ 
\\
\textbf{Acknowledgements}\\
K.S., T.P. and F.M. acknowledge the funding from European Research Council (ERC) under the European Union’s Horizon 2020 research and innovation program (MORE-TEM ERC-SYN project, Grant agreement No. 951215). J.H. and C.M. acknowledge the support from the National Natural Science Foundation of China (Grant No. 12304082) and the Natural Science Foundation of Hunan Province (2023JJ20001). K. S. would like to acknowledge the support from JST-CREST (JPMJCR20B1) and JSPS-KAKENHI (23H05443). We thank Dr. Cudazzo for providing the data for cumulant method calculation.\\ 
\\
\textbf{Author information}\\
These authors contributed equally: Jinhua Hong, Alberto Guandalini, Weibin Wu.\\
\\
Authors and Affiliations\\
College of Materials Science and Engineering, Hunan University, Changsha, China\\
Jinhua Hong, Weibin Wu, Fuwei Wu, Shulin Chen, Chao Ma\\ 
Dipartimento di Fisica, Università di Roma La Sapienza, Piazzale Aldo Moro, Roma, Italy\\
Alberto Guandalini, Francesco Mauri\\
SANKEN (The Institute of Scientific and Industrial Research), The University of Osaka, Osaka, Japan\\
Haiming Sun, Kazu Suenaga\\
University of Vienna, Faculty of Physics, Vienna, Austria\\
Thomas Pichler\\
\\
\textbf{Contributions}\\
J.H. and K.S. conceived the experiment. J.H., W.W., F.W. and H. S. carried out the h-BN sample transfer/annealing and STEM/$q$-EELS measurements and also all experimental data processing. A.G. and F.M. performed the theoretical calculations and analysis. S. C. and C. M. provided the high-quality samples and supported the TEM characterization. T. P. did the data analysis and led the discussion with all the others. J.H. and A.G. wrote the manuscript with contributions from all authors. \\
\\
\textbf{Corresponding authors}\\
Correspondence should be sent to: alberto.guandalini@uniroma1.it (A. G.), 
suenaga-kazu@sanken.osaka-u.ac.jp (K. S.)
thomas.pichler@univie.ac.at (T. P.), francesco.mauri@uniroma1.it (F. M.)
\\

\noindent\textbf{Ethics declarations}\\
\textbf{Competing interests}\\
The authors declare no competing interests.\\
\\
\textbf{Additional information}\\
Publisher’s note Springer Nature remains neutral with regard to jurisdictional claims in published maps and institutional affiliations.\\
\\
\textbf{Supplementary information}\\
Supplementary Fig. S1-S13.\\
\\
\textbf{Rights and permissions}\\
Springer Nature or its licensor holds exclusive rights to this article under a publishing agreement with the author(s) or other rightsholder(s); author self-archiving of the accepted manuscript version of this article is solely governed by the terms of such publishing agreement and applicable law. 

\bibliographystyle{apsrev4-1}
\bibliography{bibliography}

\setcounter{figure}{0}
\renewcommand{\figurename}{\textbf{Extended Data Fig.}}
\begin{figure*}[!htbp]
\includegraphics[width=1\textwidth,keepaspectratio]{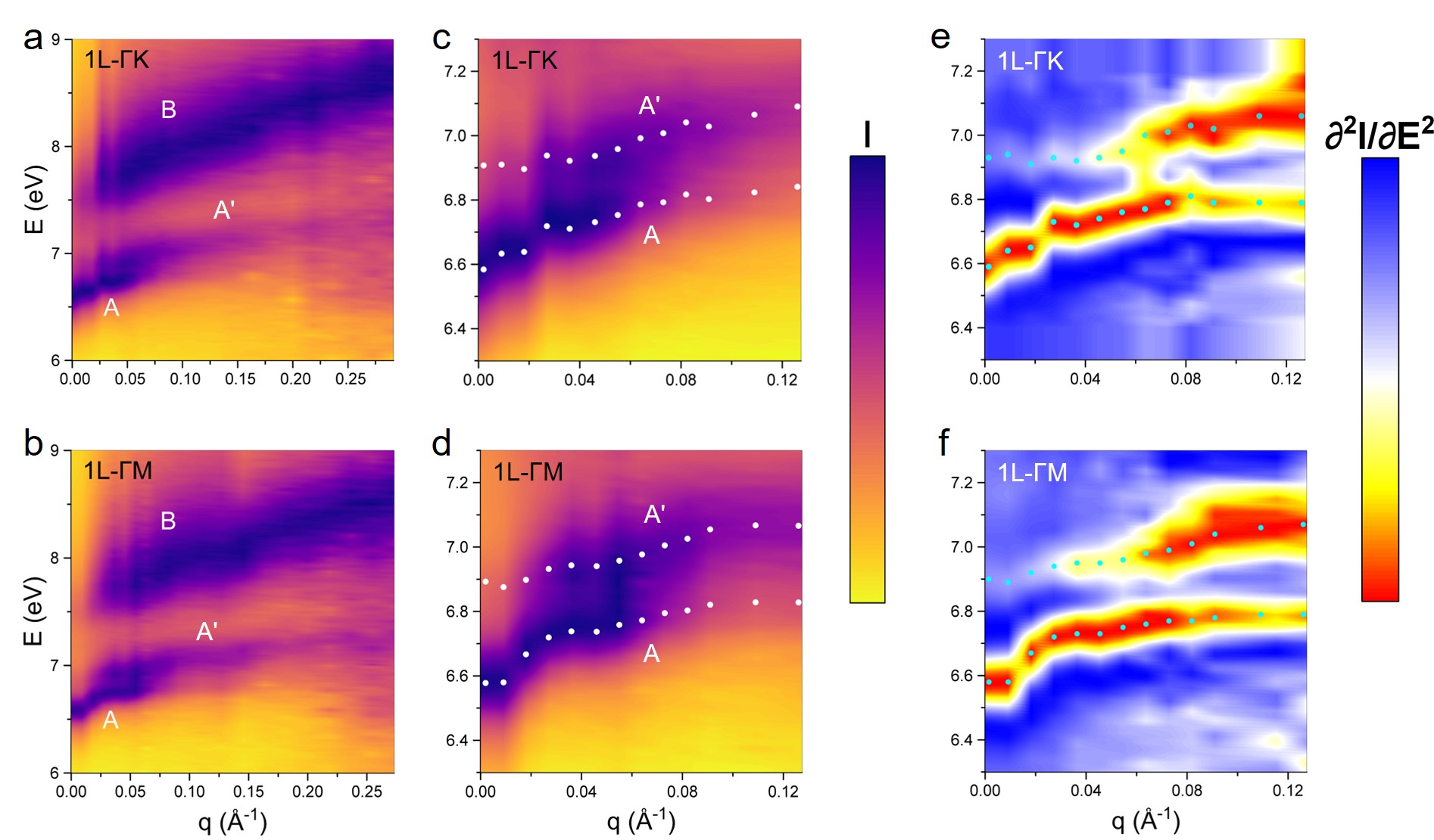}
    \caption{\textbf{Experimental $q$-E diagrams of monolayer h-BN.} a, $q$-E diagram along the $\Gamma$K direction. b, along the $\Gamma$M direction. Fine structure A, A' and B peaks are marked on the raw data without any smoothing.c-d, Zoom-in fine structure of the A and A' structure in a and b, respectively. White dots are extracted using 4LM fitting. e-f, Second derivative of the q-EEL spectrum intensity of a and b, respectively. }
    \label{fig:S17}
\end{figure*}
\newpage

\begin{figure*}[!htbp]
    \includegraphics[width=0.4\textwidth,keepaspectratio]{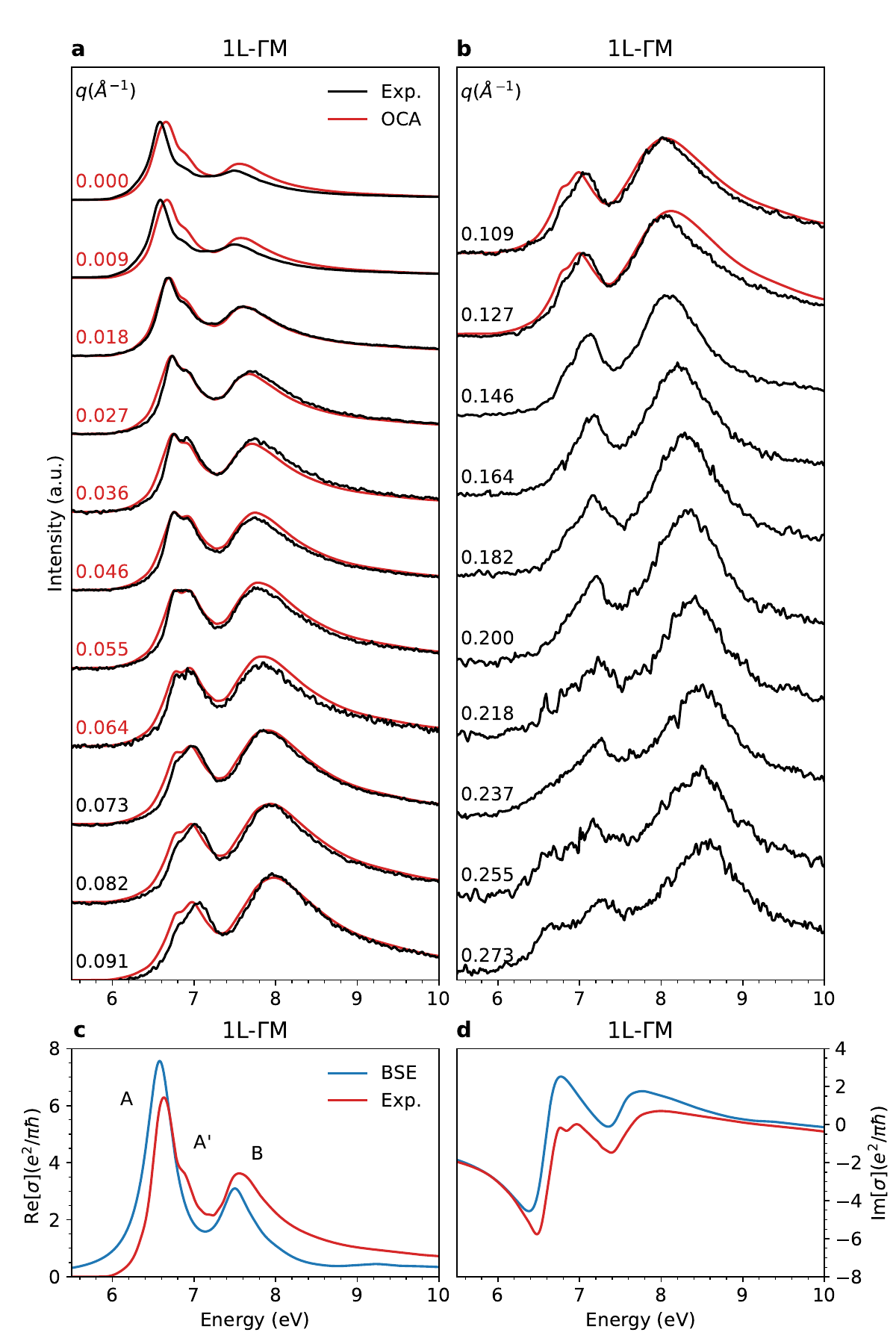}
    \includegraphics[width=0.4\textwidth,keepaspectratio]{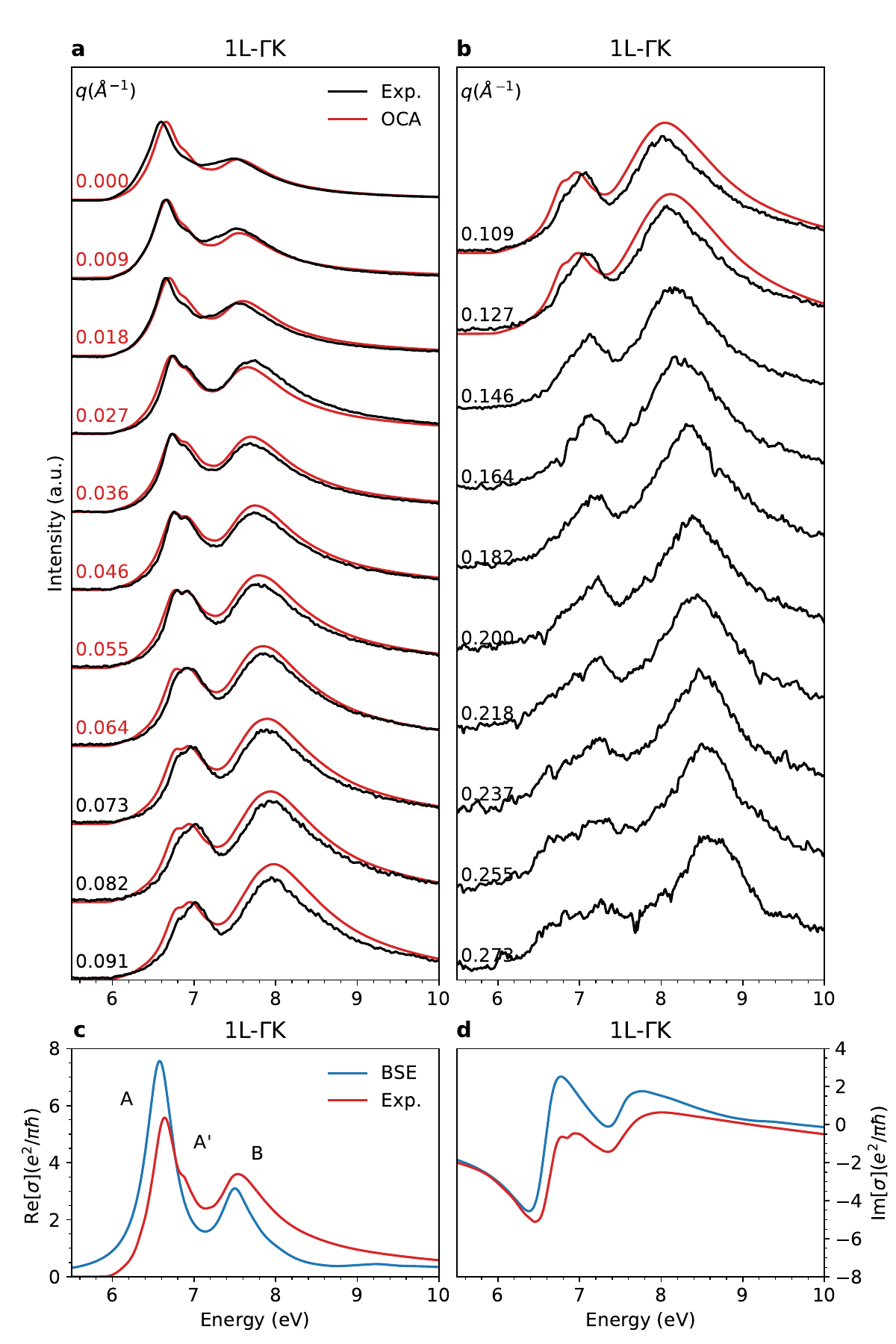}
    \includegraphics[width=0.4\textwidth,keepaspectratio]{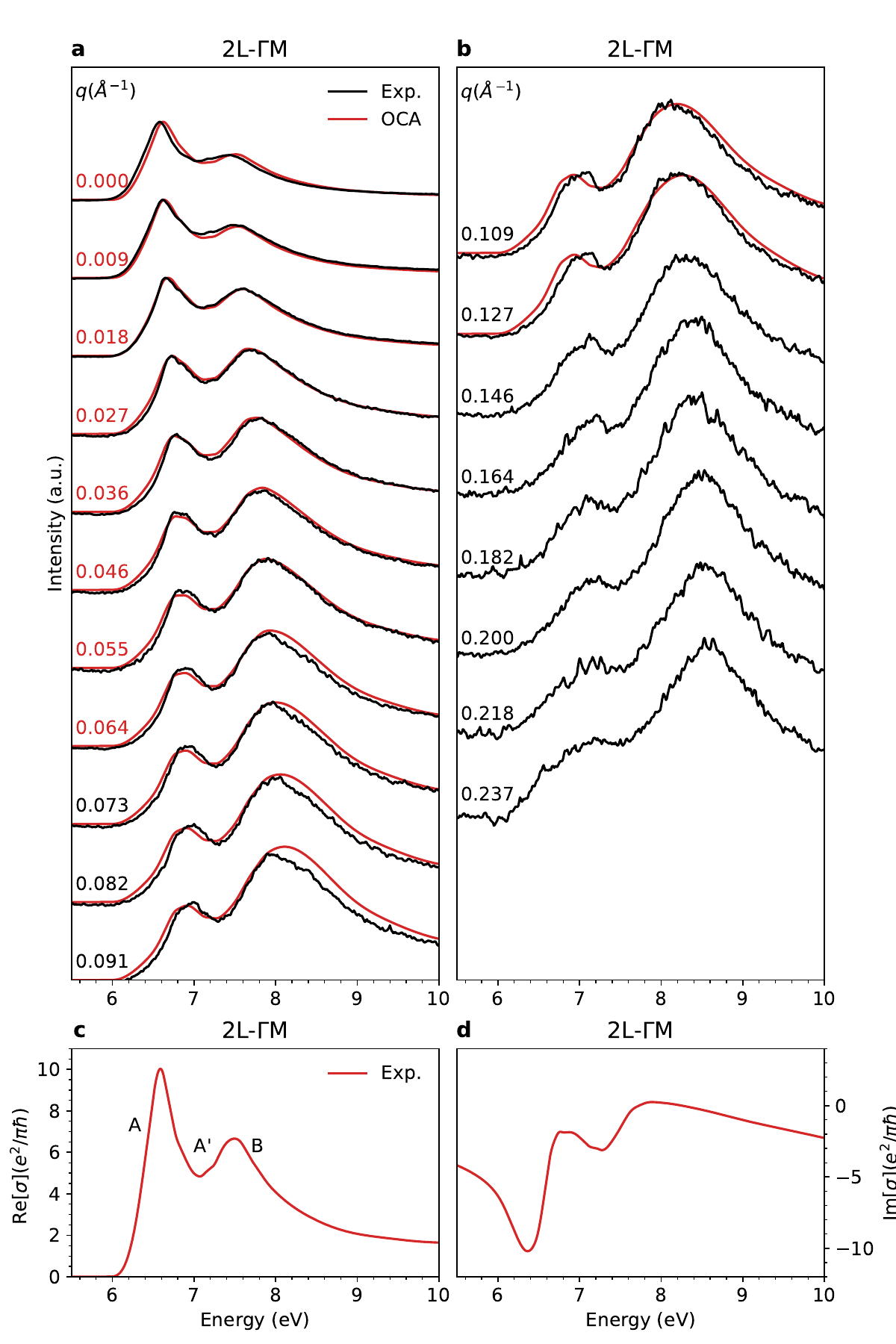}
    \includegraphics[width=0.4\textwidth,keepaspectratio]{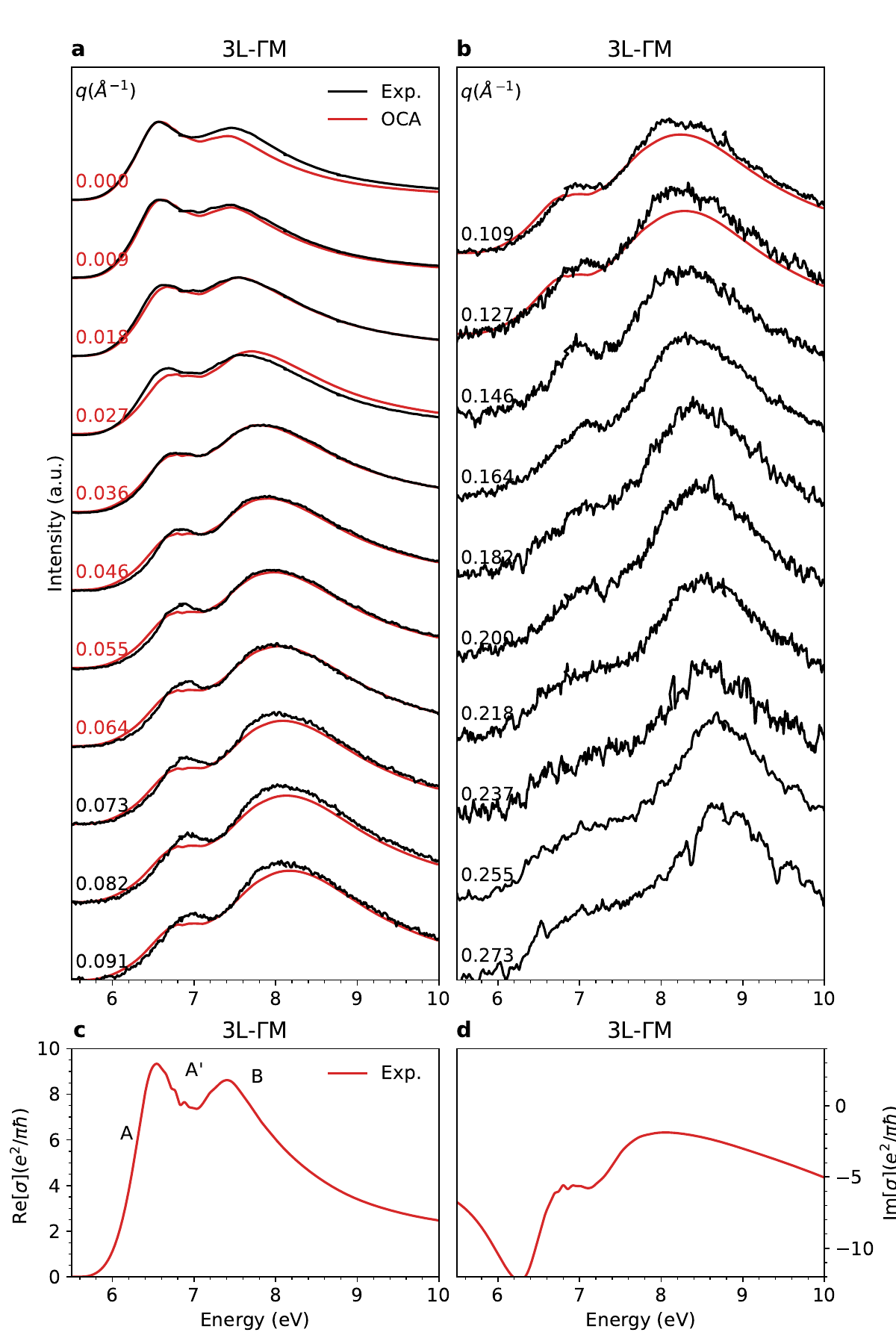}
    \caption{\textbf{Exciton peak evolution in the low q range towards the optical limit for h-BN with different thicknesses.} 
    (a)-(b): $q$-EELS experimental spectra (black) and obtained from the optical conductivity approximation [red lines of panels (a) and (b)]. Red momentum transfers correspond to the data used to extrapolate the optical conductivity. We note the optical conductivity approximation slowly deteriorates over the increasing $q$.
    Real (c) and imaginary (d) parts of the 2D optical conductivity over the number of layers in units of the quantum conductance ($G_0 = e^2/\pi\hbar$) are extracted from experiments through the optical conductivity approximation (red) and for the monolayer calculated with the Bethe-Salpeter equation at fixed nuclei (blue).
    We report here data for monolayer h-BN of the $\Gamma$M and $\Gamma$K directions, along with $\Gamma$M data of the bilayer (2$\Gamma$M) and trilayer (3$\Gamma$M)
    }
    \label{fig:spectra_extended}
\end{figure*}

\begin{figure*}[!htbp]
\includegraphics[width=0.7\textwidth, keepaspectratio]{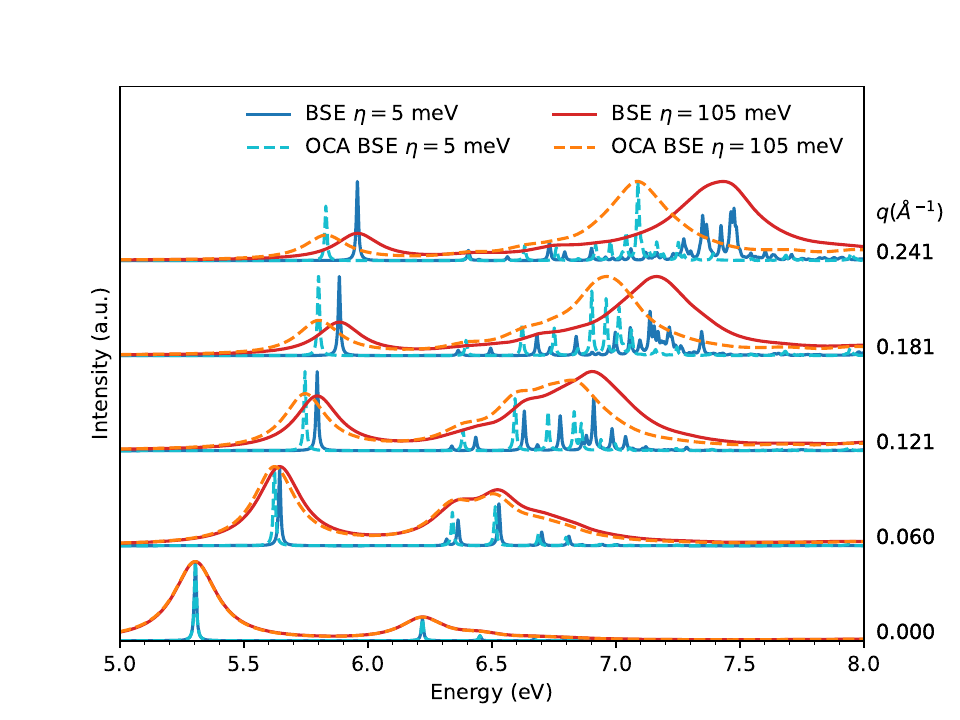}
    \caption{\textbf{Accuracy of the optical conductivity approximation.} 
    Electron energy-loss spectra obtained with the GW-BSE method with two damping factors $\eta$ (blue and light blue), compared with the same spectra obtained with the optical-conductivity approximation (red and orange).
    At $q=0$, BSE and OCA-BSE are equal by construction.
    The OCA accuracy slowly decrease by increasing $q$.
    }
    \label{fig:OCA_range}
\end{figure*}

\begin{figure*}[!htbp]
\includegraphics[width=0.7\textwidth, keepaspectratio]{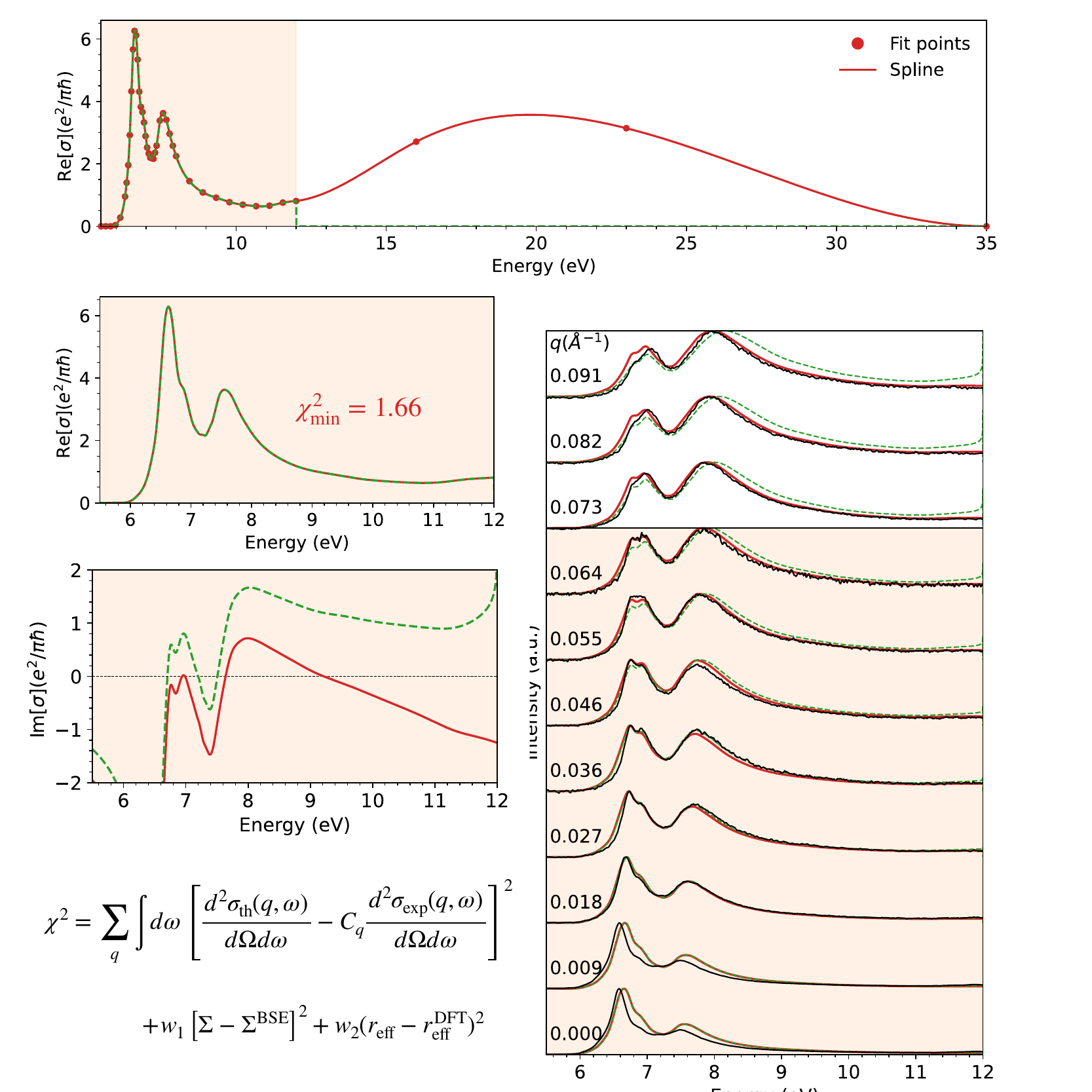}
    \caption{\textbf{Extraction of the optical conductivity $\sigma(\omega)$ from EEL spectra.}
    We extracted the optical conductivity $\sigma(\omega)$ from the EEL spectra with the following numerical procedure.
    We first define the cost function $\chi^2$ written in the figure, where $d^2\sigma_{\mathrm{th}}/d\Omega d\omega$ is calculated from the $\sigma(\omega)$ we want to extract within the optical conductivity approximation described in the method section of the main paper, $d^2\sigma_{\mathrm{exp}}/d\Omega d\omega$ are the raw EEL cross sections, $C_q$ normalization constants needed as the absolute spectra intensities are dependent on experimental conditions like exposure time etc.
    The energy integral is performed over $\hbar\omega \in [5.5,11]$ eV.
    We added in the cost function the square deviation with the BSE sum rule $\Sigma^{\mathrm{BSE}}$ in the range $E \in [0,10]$ eV and the DFT effective radius $r_{\mathrm{eff}}^{\mathrm{DFT}}$ with $w_1 = 10$ and $w_2 = 0.5$.
    We refer to the caption of Tab. I of the main paper for a definition of $\Sigma$ and $r_{\mathrm{eff}}$.
    Next, we discretize the real part of the optical conductivity Re$[\sigma(\omega)]$ over a finite number of frequencies $\lbrace\omega_i,\sigma_i \rbrace$ (top panel).
    The $\sigma(\omega)$ between the numerical points are interpolated with a third-order spline interpolation.
    Im$[\sigma(\omega)]$ is evaluated starting from Re$[\sigma(\omega)]$ with Kramers-Kronig relations.
    The error function is thus a variable of the discretized real part of the conductivity and normalization constants: $\chi^2(\lbrace\sigma_i \rbrace,\lbrace C_q \rbrace)$.
    Starting from an initial guess $C_q = 1$ $\forall \ q$ and $\sigma_i = 1$ $\forall \ i$, we minimized the cost function and the $\lbrace \sigma_i\rbrace$ are plotted with red points in the top panels, interpolated with the splines cited above.
    The fit is performed with experimental data in the range $q < 0.07$ \AA$^{-1}$, where the OCA shows to be accurate (see Fig. \ref{fig:OCA_range}).
    The correspondent EEL spectra are plotted in the right panel in red, to be compared with the experimental data in black.
    We note some $\lbrace \omega_i \rbrace$ are outside the $\omega$ integration in the cost function $\chi^2$.
    This is not problematic, as Im$[\sigma(\omega')]$, where $\omega' \in [0,11]$ eV, depend on Re$[\sigma(\omega'')]$ with $\omega'' \in [-\infty,\infty]$ through the Kramers-Kronig relations.
    We finally analyze the contribution of the high-energy part of the optical conductivity by comparing $\sigma(\omega)$ results with those obtained with $\sigma'(\omega)$, where Re$[\sigma'(\omega)] = $Re$[\sigma(\omega)]$ for $\omega < 11$ eV and Re$[\sigma'(\omega)]=0$ for $\omega > 11$ eV (green dashed curve in the top panel.)
    The correspondent spectra are plotted i the right panel with green dashed lines.
    From the similarity beween the spectra obtained from $\sigma(\omega)$ and $\sigma'(\omega)$, we conclude the high-energy contributions of the conductivity are almost negligible in this system.
    However, high-energy terms can be in principle important for other system classes.
    The same procedure, here illustrated for monolayer h-BN with $q$ oriented along the $\Gamma$M direction, has been repeated along the $\Gamma$K direction and for bi- and tri-later h-BN along the $\Gamma$M directions.}
    \label{fig:fit_explain}
\end{figure*}

\begin{figure*}[!htbp]
\includegraphics[width=0.8\textwidth,keepaspectratio]{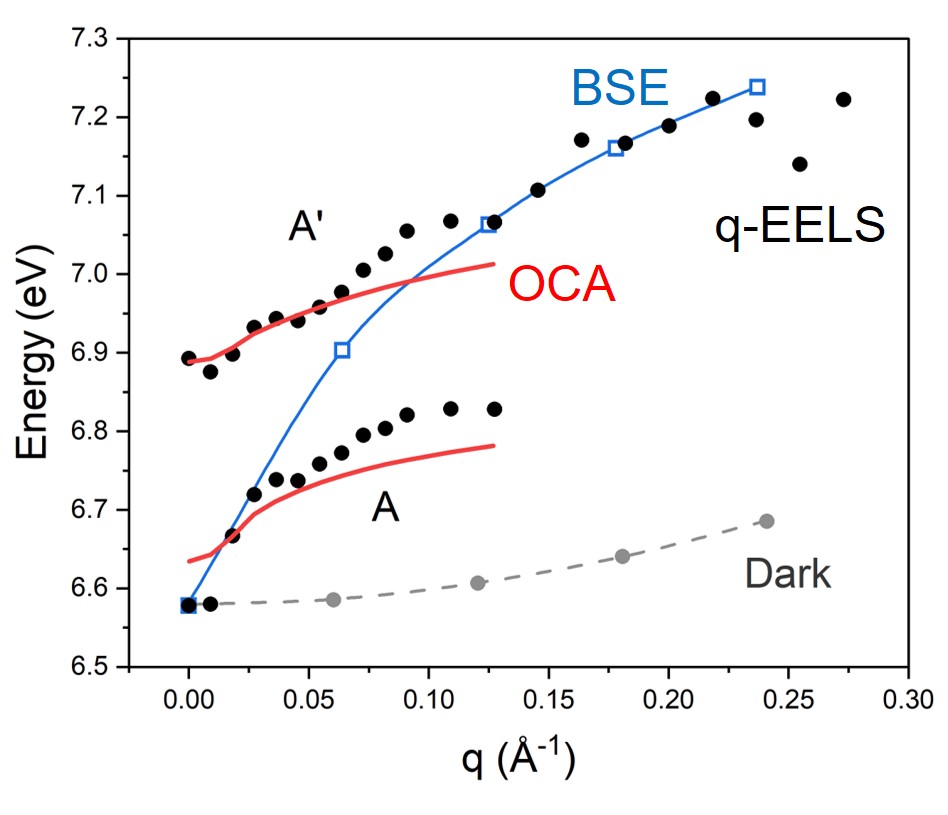}
    \caption{\textbf{Energy dispersions of monolayer h-BN in the low $q$ range along $\Gamma$M direction.} Experimental dispersions are shown by the black dots, and GW-BSE  calculated dispersions are in blue and gray. The dashed gray branch corresponds to dark exciton band which is dipole forbidden and absent in the experiment.}
    \label{fig:BSE_dark}
\end{figure*}

\begin{figure*}[!htbp]
    \includegraphics[width=0.49\textwidth, keepaspectratio]{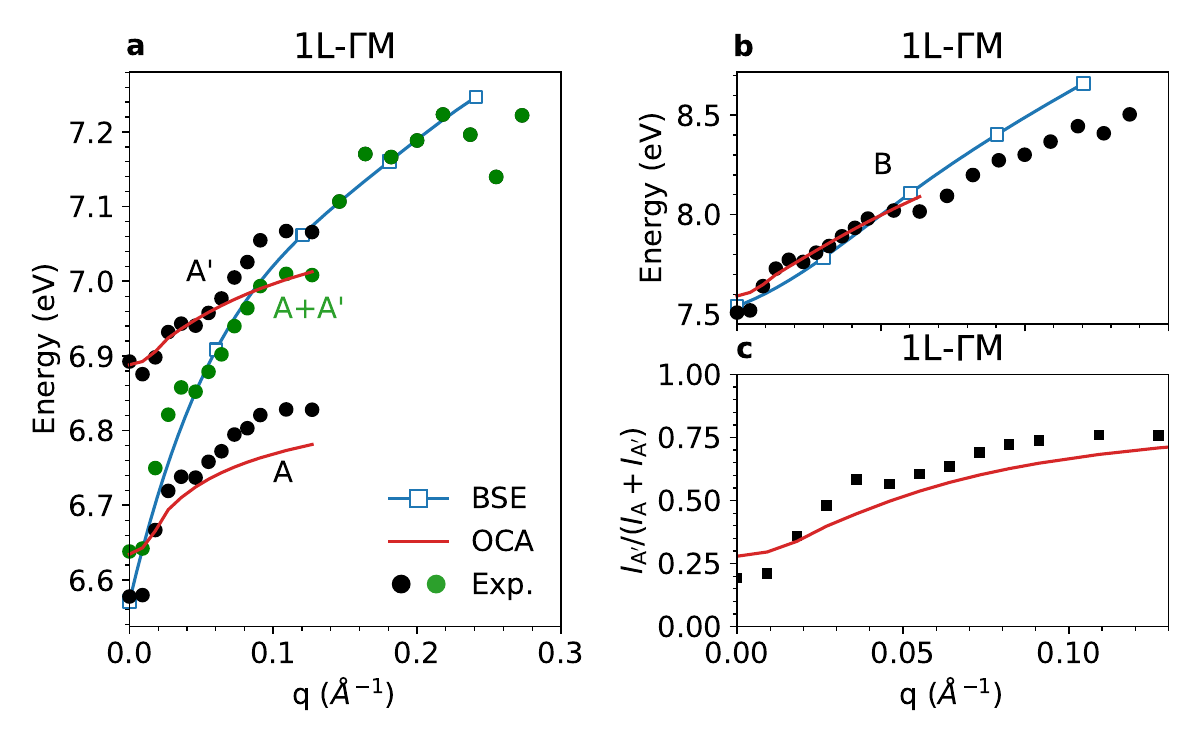}
    \includegraphics[width=0.49\textwidth, keepaspectratio]{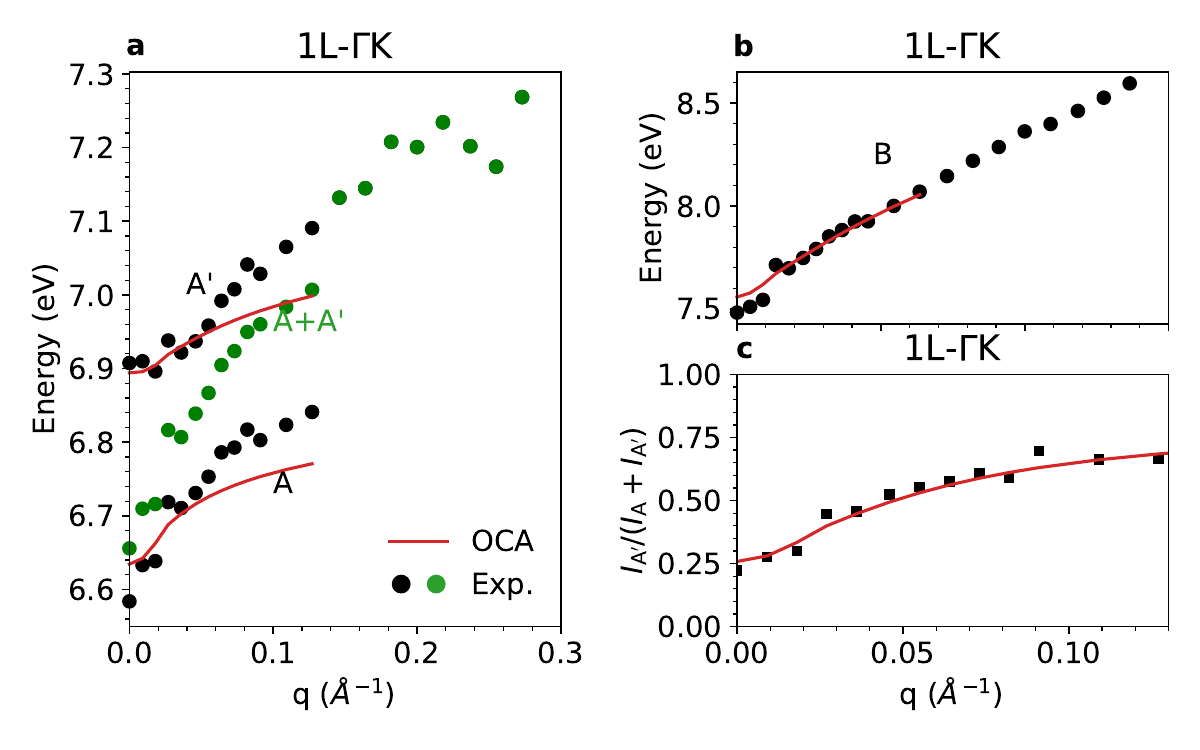}
    \includegraphics[width=0.49\textwidth, keepaspectratio]{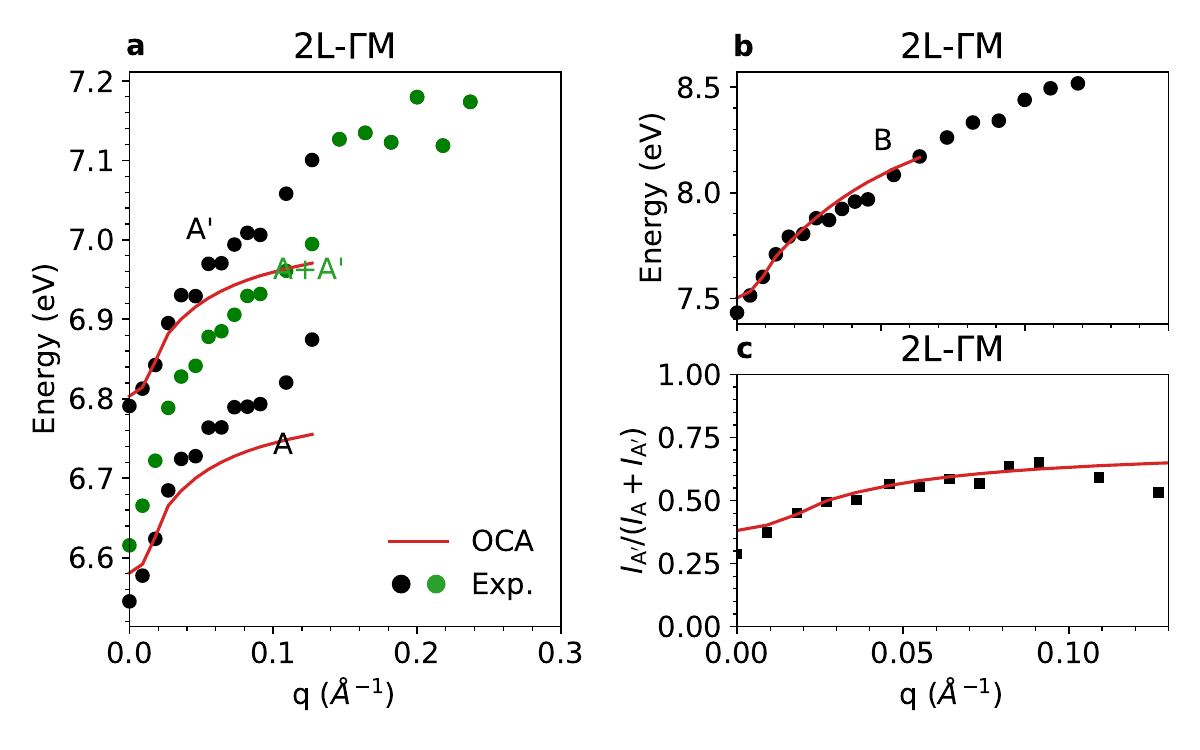}
    \includegraphics[width=0.49\textwidth, keepaspectratio]{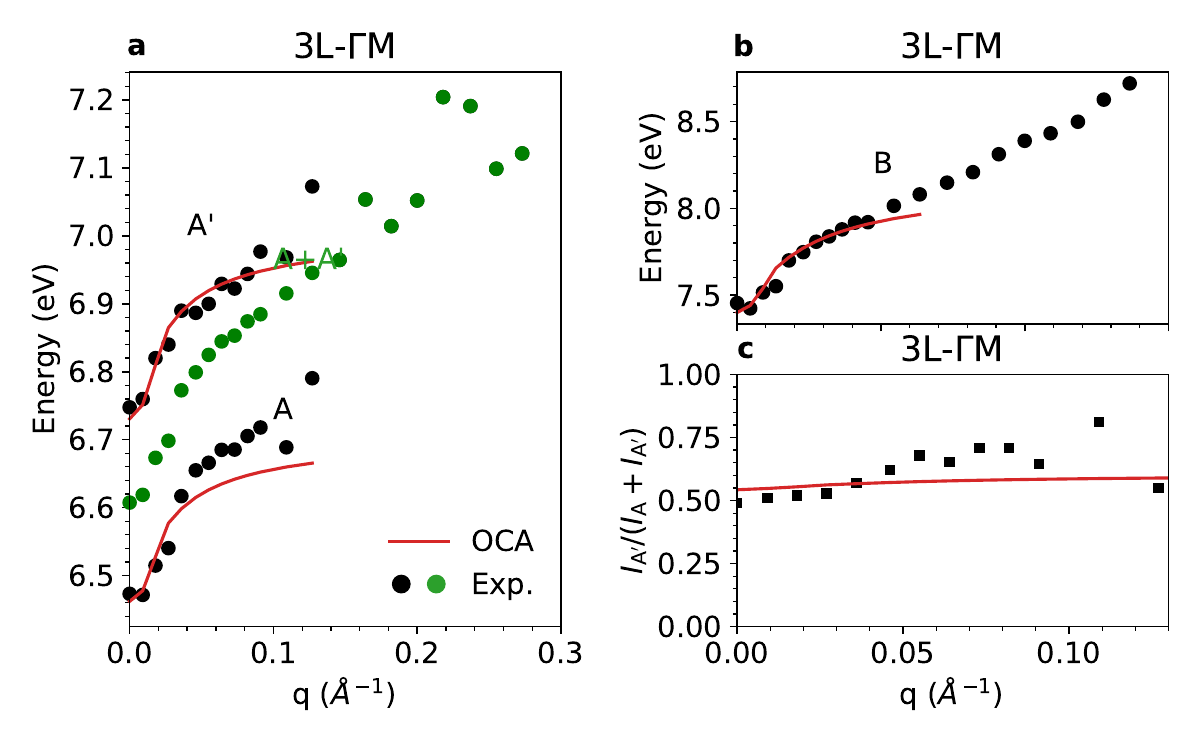}
    \caption{\textbf{Energy and oscillator strength dispersions of mono- and multi-layer h-BN} (a): Exciton dispersions of the A and A$'$ peaks in black and its weighted average (($E_AI_A+E_{A'}I_{A'})/(I_A+I_{A'})$) in green, and OCA (red) and BSE (blue) calculated dispersions. The OCA has been restricted to its range of validity up to $q_{\mathrm{max}} \approx 0.12$ \AA$^{-1}$
    (b): The dispersion of the B peak obtained by experiment (black), OCA (red) and BSE (blue).
    (c): Fraction of the intensity of A$'$ peak with respect to the A+A$'$ peaks.
    We report here data for monolayer h-BN of the $\Gamma$M and $\Gamma$K directions, along with $\Gamma$M data of the bilayer (2$\Gamma$M) and trilayer (3$\Gamma$M)
    }
    \label{fig:dispersion_extended}
\end{figure*}

\begin{figure*}[!htbp]
\includegraphics[width=1\textwidth,keepaspectratio]{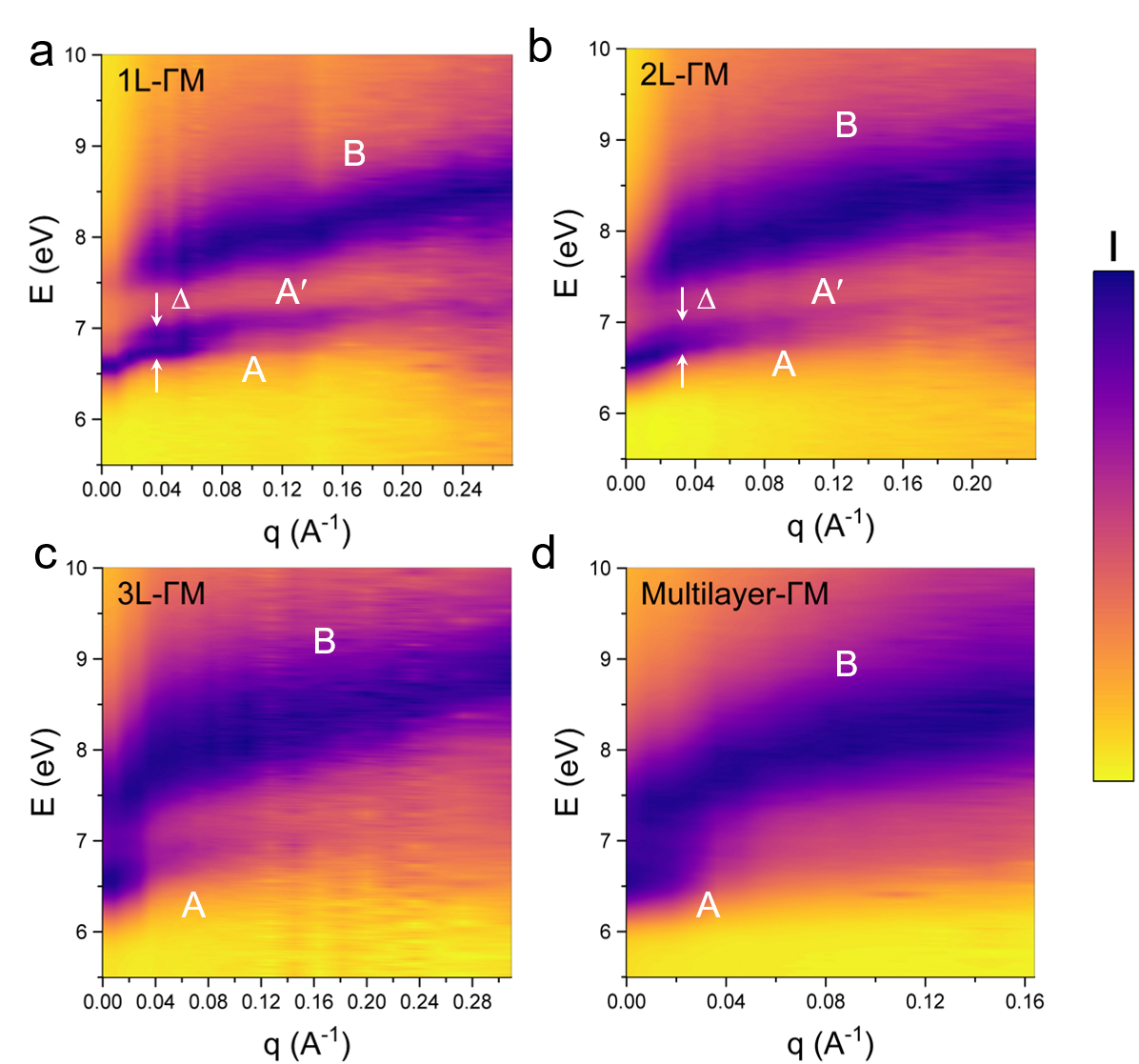}
    \caption{\textbf{$q$-E diagram of h-BN with different thicknesses along the $\Gamma$M direction.} a, monolayer h-BN. b, bilayer h-BN. c, tri-layer h-BN. d, multilayer h-BN. The A exciton band shows an obvious splitting in the monolayer and get weakened with the increasing number of layers. }
    \label{fig:S14}
\end{figure*}

\begin{figure*}[!htbp]
\includegraphics[width=1\textwidth,keepaspectratio]{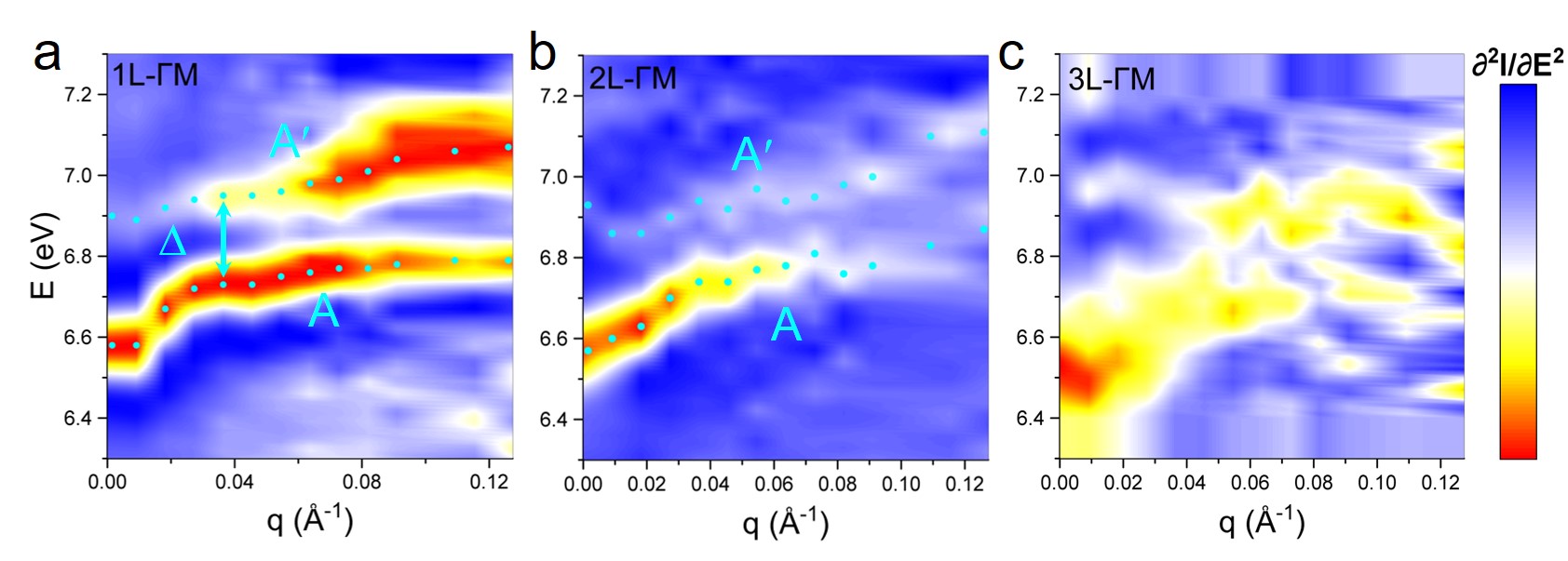}
    \caption{\textbf{$q$-E diagram of h-BN with different thicknesses along the $\Gamma$M direction in the form of second derivative of the spectrum intensity.} a, monolayer h-BN. b, bilayer h-BN. c, tri-layer h-BN. The A-A$'$ splitting gap is most prominent in monolayer, and decrease in bilayer, and then becomes invisible as the thickness further increases. }    \label{fig:second_derivative}
\end{figure*}

\begin{figure*}[!htbp]
\includegraphics[width=1\textwidth,keepaspectratio]{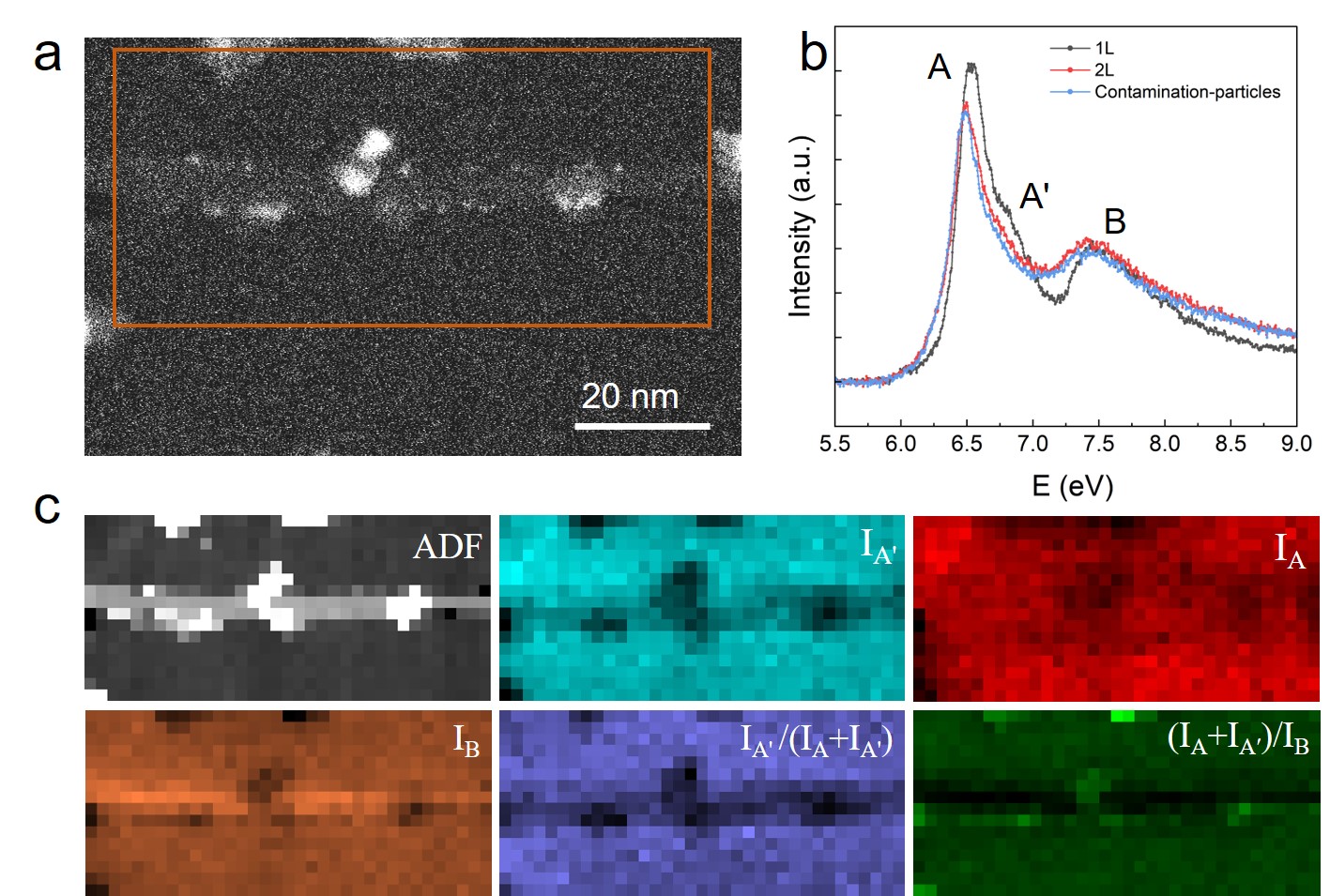}
    \caption{\textbf{STEM-EELS mapping of the exciton fine structure.} (a): ADF-STEM image of a nanoribbon on top of monolayer h-BN forming a monolayer-bilayer-monolayer transition. The orange rectangle mark the area for spectrum imaging in c. (b): The spectrum fine structure of clean monolayer and bilayer and contaminated nanoparticles in the orange rectangle in a. Note the A$'$ structure is prominent in monolayer. (c): The intensity distribution of the A-A$'$ peaks. The A$'$ structure is manifested in the clean monolayer region and becomes invisible in bilayer and defective regions. Compared with the A exciton, the A$'$ is much more sensitive to contamination and thickness, indicating it arises from intrinsic excitation properties.}
    \label{fig:monomap}
\end{figure*}

\end{document}